\documentclass[12pt]{article}
\usepackage{amssymb}
\usepackage{amsfonts}
\usepackage{amsmath,amsthm}
\usepackage{graphicx,epsfig,  color}

\usepackage{subfigure}
\def\be{\begin{equation}}
\def\ee{\end{equation}}
\begin{document}
\begin{center}
{{\bf {Wave optics and weak gravitational lensing of lights from
spherically symmetric static scalar vector Brans Dicke
 black holes in presence of the cosmological
constant}}
 \vskip 0.5 cm
  {{Hossein Ghaffarnejad}}\footnote{E-mail address:
hghafarnejad@semnan.ac.ir. }\vskip 0.1 cm \textit{  Faculty  of
Physics, Semnan University, 35131-19111, Semnan, Iran}}
\end{center}
%\vspace{0.1cm}
 \begin{abstract}
This paper has three parts. In first step we use modified scalar
tensor vector Brans Dicke gravity \cite{GH0} to obtain metric
solution of a spherically symmetric static black hole via
perturbation method which asymptotically behaves as modified
Schwarzschild de Sitter black hole in weak field limits.
Corrections on the line element with respect to the point mass
Schwarzschild de Sitter black hole are two parts: power law
function for black hole mass instead of point mass and additional
logarithmic metric potential which they are produced from effects
of interacting timelike dynamical vector fields. Second part of
the paper is dedicated to review Fresnel-Kirchhoff diffraction
theory for massless scalar field instead of the vector
electromagnetic waves. In third part of the work we consider the
obtained black hole metric to be lens and we study weak
gravitational lensing of moving light originated from point star.
Production of stationary images is investigated via both
interference of waves and geometric optics approaches. At last we
compare results of these approaches and give some outlooks.
\end{abstract}
\section{Introduction}
Gravitational lensing is one of the predictions of Einstein`s
general relativity theory in which the deflection of light rays
coming from a distant star and passing of near the sun is solved
by the lens equation. This is done in weak gravitational field
with thin lens approximation \cite{SE}. This method of
gravitational lensing is called as geometrical optics
approximation in which the light ray is traced. This method is
used by scientific authors for black holes and cosmological
models. There are now many articles in the literature where the
authors proposed several types of lens equations for tracing the
bending light for which one can see for instance \cite{HN,HM,HG}
and references therein. Many samples of images caused by
gravitational lensing have been obtained observationally.\\
 In fact we call geometrical optics for a path of light ray derived from the high frequency Maxwell electromagnetic waves where the
 wavelength is so much smaller than the size of lens objects.
 If this is not happened, then the wave effects of the light rays namely diffraction and interference patterns, become important
 and so we must take into account the wave properties of the light rays.
  For
instance one can see \cite{TT,CB,NM} to study wave effects on the
amplification factor of wave intensity and \cite{HF,MM} for direct
detection  of the black holes via imagining shadows. We know that
the apparent angular size of the black hole shadows are small and
so their detectability depends on angular resolution of a used
telescope.
 Resolution of a telescope is determined by diffraction limit of image formation system where wave effects on image
 have more important for successful detection of black holes shadows. Scattering of waves by a Schwarzschild black hole is
 studied \cite{KI} to obtain images of the black hole from scattered wave data for particular scattering angles.
  Diffraction of a plan Electromagnetic wave by a Schwarzschild black hole was studied \cite{EH,EL,HH} to obtain conditions
  in which the geometrical limits of the optics become dominant.
  Authors in \cite{SD} examined diffraction effects of waves on gravitational lensing instead of the geometrical optics approximation where mass of the lens objects is smaller than the sun mass.
   They obtained a pattern of interference fringes for which diffraction phenomena, intensity of images on the optical axis with the lowest mass the significant brightening occurs while, in the geometrical optics approximation it is infinite.
    They showed that the diffraction of the waves for low mass stars such as planets give out a suitable way to probe the unknown dark matter.
     One can obtain wave effects of gravitational waves in ref. \cite{RT} which are important in the gravitational lensing when their wavelength is greater than the Schwarzschild radius of the lens.
     Authors of the latter work showed that with how much accurate can extract the mass of the lens and the source position from signal of the lens.
     \\
According to the work \cite{YN} presented by Yasusada Nambu we
want to use wave behavior of the light ray in the gravitational
lensing of our obtained black hole metric as lens. In fact we seek
gravitational effects of time like vector field on formation of
images in the weak gravitational lensing for stationary images via
approaches of wave optics and geometric optics. We restrict this
work to weak gravitational lensing  in which the assumption is
that the moving lights waves are far from the center of the black
hole and so effects of the horizon and the photon sphere can be
negligible.
 For this purpose we apply the Fresnel-Kirchhoff diffraction
theory of image formation in wave optics. Here we say a little
about importance of the scalar vector Brans Dicke gravity theory
which we used in this work. To give an answer to open problem in
the literature where how can changes signature of the background
metric from Lorentzian $(-+++)$ to Euclidean $(++++)$ dynamically?
which can be responsive for state of before the Big Bang phase of
the expanding universe (see for instance \cite{GH2}), more
alternative scalar vector tensor gravities called as bimetric
gravity models are presented. They can be followed at references
of the work \cite{GH0} which we used here. In the latter model the
well known Brans Dicke scalar tensor gravity is changed to an
alternative form by metric transformation $g_{\mu\nu}\to
g_{\mu\nu}+2N_{\mu}N_{\nu}$ in which $N_{\mu}$ is a time like
dynamical vector field. The main role of this vector field is
describe how should be dynamics of metric signature transition of
a curved space time metric.
 Thereinafter we studied some applications of this alternative gravity model: For instance study of signature transition
 of Robertson Walker space time with canonical quantum cosmology approach \cite{GH2}.  Study of canonical quantization of
 isotropic\cite{GH3}  and anisotropic \cite{GH4} cosmology. Study of dynamical stability of $\Lambda CDM$ model for isotropic
  Robertson walker \cite{GH5} and anisotropic Bianchi I \cite{GH6} cosmology. Study of vector field corrections on the galaxy rotation
  curves \cite{AF}.\\
As a different metric solution of this model with black hole
topology we solved linear order Einstein metric equation via
perturbation method and obtained a spherically symmetric static
black hole which asymptotically behaves similar to a modified
Schwarzschild de Sitter black hole. In this metric solution
dynamical timelike vector field produces a logarithmic metric
potential and also mass function for the black hole instead of a
 point mass. In the present work we want to study effects of these two additional parts on the interference
 fringes and image formation of the weak gravitational lensing. Layout of the paper is as follows. \\
In section 2 we define scalar vector tensor Barns Dicke gravity
briefly and then solve metric field equation to obtain
 a black hole metric solution.
According to the work \cite{RT,YN} presented by Nambu et al we
review in section 3 general formalism of the Fresnel-Kirchhoff
diffraction theory for scattering scalar waves and image
formation. In section 4 we calculate explicit form of the lens
potential for our obtained
 black hole metric solution. Also we calculate image amplification factor and obtain positions of stationary images
 in both of approaches namely geometric optics and wave optics for different numeric values of the vector field parameter.
 Section 5 denotes to outlook of the work.
 \section{Scalar vector tensor Brans Dicke black hole}
 Let us we start with the following gravity model \cite{GH0}
 \begin{equation}\label{Itot} I_{total}=I_{BD}+I_{N}\end{equation}
where
\begin{equation}I_{BD}=\frac{1}{16\pi}\int dx^4 \sqrt{g}\{\phi R-\frac{\omega}{\phi}g^{\mu\nu}
\partial_{\mu}\phi\partial_{\nu}\phi\}\end{equation}is the Jordan Brans Dicke scalar tensor gravity \cite{BD} with dimensionless parameter $\omega$ and
\begin{equation}I_{N}=\frac{1}{1\pi}\int dx^4 \tilde{\zeta}(x^\nu)(g^{\mu\nu}N_{\mu}N_{\nu}+1)+U(\phi,N_{\nu})+2\phi F_{\mu\nu}F^{\mu\nu}
\end{equation}$$-\phi N_{\mu}N^{\nu}(2F^{\mu\lambda}\Omega_{\nu\lambda}
+F^{\mu\lambda}F_{\nu\lambda}+\Omega^{\mu\lambda}\Omega_{\nu\lambda}-2R^{\mu}_{\nu}+\frac{2\omega}{\phi^2}\partial^{\mu}\phi
\partial_{\nu}\phi)\}$$ with \begin{equation}F_{\mu\nu}=2(\nabla_{\mu}N_{\nu}-\nabla_{\nu}N_{\mu}), ~~~~\Omega_{\mu\nu}=2(\nabla_{\mu}N_{\nu}+
\nabla_{\nu}N_{\mu}).\end{equation} Without $\tilde{\zeta}$ and
$U$ the latter action $I_{N}$ is generated from $I_{BD}$ under the
particular metric transformation
$g_{\mu\nu}\to2N_{\mu}N_{\nu}+g_{\mu\nu}.$
 The undetermined Euler Lagrange coefficient $\tilde{\zeta}(x^{\nu})$ shows that the dynamical vector field $N_{\nu}$
  treats as time like and so can be interpreted as four velocity of preferred reference
  frame. We now want to study the above theory for a spherically symmetric static space time with a black hole topology which in general
  form is given by the following line element.
   \begin{equation}\label{line1}ds^2=-e^{2\epsilon \alpha(r)}dt^2+e^{2\epsilon \beta(r)}\{dr^2+r^2d\theta^2+r^2\sin^2\theta d\varphi^2\}
 \end{equation} where we bring the parameter $\epsilon$ to use an order parameter if when we apply to solve the metric
 field equations by perturbation method in what follows. By
 varying $I_N$ with respect to $\tilde{\zeta(n^\nu)}$ we obtain
 $g^{\mu\nu}N_{\mu}N_{\nu}=-1$ which for the above line element
 reaches \begin{equation}\label{Nt}N_{t}(r)=\tilde{\xi} e^{\epsilon\alpha(r)},~~~N_{r}(r)=\sinh\tilde{\xi} e^{2\epsilon \beta(r)},~~~N_{\theta}=0=N_{\varphi}
 .\end{equation}
This particular choice of vector field shows radially accelerated
local preferred reference frame.  The $\tilde{\xi}$ parameter with
real numeric values determines in fact polar direction of the
vector field $N_{\mu}$ in the spacetime. One can substitute the
line element (\ref{line1}) into the action functional (\ref{Itot})
to show
\begin{equation}\label{action2} I_{total}=\frac{1}{4G}\int dt\int dr\{(1+2\sigma
)r^2e^{\epsilon(\alpha-\beta+\psi)}(\alpha^{\prime\prime}+2\beta^{\prime\prime})
\end{equation}$$+r^2e^{\epsilon(\alpha+\beta)}U(\epsilon\psi,\sigma,\alpha,\beta)\}+\frac{\epsilon}{4G}\int dt\int dre^{\epsilon(\alpha-\beta+\psi)}\{2r^2\beta^{\prime2}+(1-2\sigma)r^2\alpha^\prime\beta^\prime
$$$$+(1+2\sigma)r^2\alpha^{\prime2}+4r\beta^\prime+4(1-\sigma)r\alpha^\prime-2(1+\sigma)r^2\alpha^{\prime\prime}\}+\frac{\epsilon^2}{4G}\int dt\int dr
$$$$+\omega e^{\epsilon(\alpha+\psi)}(e^{-\epsilon\beta}-2\sigma\omega
e^{\epsilon\beta})r^2\psi^{\prime2}-16(1+\sigma)^2e^{\epsilon(3\alpha-\beta+\psi)}r^2\alpha^{\prime2}$$$$+8\sqrt{\sigma}(1+\sigma)^\frac{3}{2}e^{\epsilon\psi}r^2\alpha^{\prime2}
+4\sqrt{1+\sigma}\sigma^\frac{3}{2}e^{\epsilon(2\alpha-2\beta+\psi)}r^2\alpha^\prime\beta^\prime$$
$$-4(1+\sigma)e^{\epsilon(\alpha-\beta+\psi)}r^2\{(5+\sigma\sqrt{1+\sigma})\alpha^{\prime2}+(1+4\sigma\sqrt{1+\sigma})\alpha^\prime\beta^\prime\}$$
where we defined \begin{equation}\label{sigma}
\sigma=\sinh^2\tilde{\xi},~~~~\phi(r)=\frac{e^{\epsilon\psi(r)}}{G}.\end{equation}
 Applying Taylor series
expansion of the fields
$\phi(r),e^{\epsilon\alpha(r)},e^{\epsilon\beta(r)}$ near the
order parameter $0<\epsilon<1$ as
\begin{equation}\label{exp}\{e^{\epsilon\psi(r)},e^{2\epsilon\alpha(r)},e^{2\epsilon\beta(r)}\}\approx1+\epsilon\{\psi(r),2\alpha(r),2\beta(r)\}+O(\epsilon^2),
\end{equation}
and
\begin{equation}U(\epsilon\psi,\sigma,\alpha,\beta)\approx U_0(\sigma,\alpha,\beta)+\epsilon U_1(\sigma,\psi,\alpha,\beta)+O(\epsilon^2)\end{equation}
one can show that the lagrangian density in the action functional
(\ref{action2}) reduces to the following series expansion.
\begin{equation}\label{L}\mathcal{L}\approx
(1+2\sigma)r^2(\alpha^{\prime\prime}+2\beta^{\prime\prime})+r^2U_0(\mu,\alpha,\beta)\end{equation}$$+
\epsilon\{2r^2\beta^{\prime2}+(1-2\sigma)r^2\alpha^{\prime}\beta^\prime+(1+2\sigma)r^2\alpha^{\prime2}
+4(1-\sigma)r\alpha^\prime-2(1+\sigma)r^2\alpha^{\prime\prime}$$$$
+r^2(\alpha-\beta+\psi)(\alpha^{\prime\prime}+2\beta^{\prime\prime})+r^2(\alpha+\beta)U_0+r^2U_1\}$$$$
+\epsilon^2\{\omega
(1-2\sigma\omega)r^2\psi^{\prime2}-16(1+\sigma)^2r^2\alpha^{\prime2}+8\sqrt{\sigma}(1+\sigma)^\frac{3}{2}r^2\alpha^{\prime2}+r^2(\alpha+\beta)U_1
$$$$
+4\sqrt{1+\sigma}\sigma^\frac{3}{2}r^2\alpha^\prime\beta^\prime-4(1+\sigma)r^2[(5+\sigma\sqrt{1+\sigma})\alpha^{\prime2}
+(1+4\sigma\sqrt{1+\sigma})\alpha^\prime\beta^\prime]\}$$$$+O(\epsilon^3).$$
Substituting linear potentials
\begin{equation}U_0=a\alpha+b\beta,~~~U_1=p\alpha+q\beta+s\psi\end{equation} and the lagrangian density (\ref{L})
into the Euler Lagrange equation
\begin{equation}\frac{\partial\mathcal{L}}{\partial \chi_i}-\frac{d}{dr}\bigg(\frac{\partial\mathcal{L}}{\partial\chi_i^\prime}\bigg)
+\frac{d^2}{dr^2}\bigg(\frac{\partial\mathcal{L}}{\partial\chi_i^{\prime\prime}}
\bigg)=0\end{equation} where $a,b,p,q,s$ are constant
 parameters and $i=\{1,2,3\}$ and
$\chi_i=\{\psi,\alpha,\beta\},$ we obtain linear second order
ordinary differential equations for the fields
$\psi(r),\alpha(r),\beta(r)$ respectively as follows.
\begin{equation}\label{psi}r^2[\alpha^{\prime\prime}+2\beta^{\prime\prime}-2\epsilon\omega(1-2\sigma\omega)\psi^{\prime\prime}]-
4\omega\epsilon(1-2\sigma\omega)r\psi^\prime
\end{equation}$$-2\omega\epsilon(1-2\sigma\omega)\psi+r^2s\epsilon(\alpha+\beta)+sr^2=0,$$
\begin{equation}\label{alpha}r^2(A_1\alpha^{\prime\prime}+A_2\beta^{\prime\prime}+\psi^{\prime\prime})+
r(A_3\alpha^\prime+A_4\beta^\prime+2\psi^\prime)+2[1+r^2(a+\epsilon
p)]\alpha
\end{equation}$$+2[-1+r^2(a+\epsilon p)]\beta+2(2+\epsilon
sr^2)\psi+\frac{2(1+2\sigma)-8\epsilon\sigma+r^2(a+\epsilon
p)}{\epsilon}=0$$ and
 \begin{equation}\label{beta}r^2(A_5\alpha^{\prime\prime}
-8\beta^{\prime\prime}+2\psi^{\prime\prime})+r(A_6\alpha^\prime-12\beta^\prime+4\psi^\prime)
+[4+r^2(a+b+\epsilon q)]\alpha\end{equation}$$+[-4+r^2(2b+\epsilon
q)]\beta+4\psi+\frac{4(1+2\sigma)+r^2(b+\epsilon q)}{\epsilon}=0$$
where we defined
\begin{equation}\label{A1}A_1=4(1+\sigma)[1-2\epsilon(9+4\sigma)+2\epsilon\sqrt{\sigma(1+\sigma)}(2-\sqrt{\sigma})]\end{equation}
\begin{equation}\label{A2}A_2=2\sigma+4\epsilon\sqrt{1+\sigma}\sigma^\frac{3}{2}-4\epsilon(1+\sigma)(1+4\sigma\sqrt{1+\sigma})\end{equation}
\begin{equation}\label{A3}A_3=6+8\sigma-64\epsilon(1+\sigma)^2+32\epsilon\sqrt{\sigma}(1+\sigma)^\frac{3}{2}-16\epsilon(1+\sigma)(5+\sigma\sqrt{1+\sigma
})\end{equation}
\begin{equation}\label{A4}A_4=-4+4\sigma+8\epsilon\sqrt{1+\sigma}\sigma^\frac{3}{2}-8\epsilon(1+\sigma)(1+4\sigma\sqrt{1+\sigma})\end{equation}
\begin{equation}\label{A5}A_5=2\sigma-4\epsilon\sqrt{1+\sigma}\sigma^\frac{3}{2}+4\epsilon(1+\sigma)(1+4\sigma\sqrt{1+\sigma})\end{equation}
\begin{equation}\label{A6}A_6=2+4\sigma-8\epsilon\sqrt{1+\sigma}\sigma^\frac{3}{2}+8\epsilon(1+\sigma)(1+4\sigma\sqrt{1+\sigma})\end{equation}
 With initial condition \begin{equation}\label{s} s=0,~~~~~\omega(\sigma)=\frac{1}{2\sigma}\end{equation}
the equation (\ref{psi}) reads
\begin{equation}\label{C}\alpha(r)+2\beta(r)=C_1r+C_2\end{equation} in which $C_{1,2}$ are integral constants. Substituting (\ref{C}) into the
equation $(\ref{beta})-2\times(\ref{alpha})$ and setting
\begin{equation}q=2p,~~~C_1=C_2=0\end{equation}
\begin{equation}\label{s0}
s_0=2\bigg\{\frac{1-8\epsilon[7+2\sigma+\sqrt{\sigma(1+\sigma)}(\sqrt{\sigma}-1)]}{1-4\epsilon[9+4\sigma-\sqrt{\sigma(1+\sigma)}
(2-\sqrt{\sigma})]}\bigg\}\end{equation}
\begin{equation}s_1=\frac{1}{8\epsilon}\bigg(\frac{b-2a}{(1+\sigma)[1-4\epsilon[9+4\sigma-\sqrt{\sigma(1+\sigma)}
(2-\sqrt{\sigma})]}\bigg)\end{equation} and
\begin{equation}s_2=\frac{\sigma}{(1+\sigma)(1-4\epsilon[9+4\sigma-\sqrt{\sigma(1+\sigma)}
(2-\sqrt{\sigma})])}
\end{equation}we obtain \begin{equation}r^2\beta^{\prime\prime}+s_0r\beta^\prime+s_1r^2+s_2=0\end{equation}
which has solution as follows.
\begin{equation}\label{betac}\beta(r)=C_3-\frac{C_4}{(s_0-1)}\frac{1}{r^{s_0-1}}-\frac{s_1}{2(1+s_0)}r^2-\frac{s_2\ln r}{(s_0-1)}\end{equation}
in which $C_{3,4}$ are integral constants. If we remember the
Schwarzschild de Sitter black hole metric $ds^2=-(1-2M/r-\Lambda
r^2/3)dt^2+(1-2M/r-\Lambda
r^2/3)^{-1}dr^2+r^2d\theta^2+r^2\sin^2\theta d\varphi^2$
\cite{GibH} which in weak field limits become
$ds^2\approx-(1-2M/r-\Lambda r^2/3)dt^2+(1+2M/r+\Lambda
r^2/3)dr^2+r^2d\theta^2+r^2\sin^2\theta d\varphi^2$ we can set
constants of the solution (\ref{betac}) as
\begin{equation} \frac{2\epsilon C_4}{s_0-1}=-(2M)^{s_0-1},~~~2\epsilon C_3=\frac{2\epsilon s_2}{s_0-1}\ln 2M,~~~~\frac{\epsilon s_1}{1+s_0}
=-\frac{\Lambda}{3},~~~\frac{2\epsilon
s_2}{s_0-1}=\mu.\end{equation}  In this case the equation
(\ref{betac}) can be rewritten as follows.
\begin{equation}\label{epsbet}2\epsilon\beta(r)=\bigg(\frac{2M}{r}\bigg)^{s_0-1}+\frac{\Lambda
r^2}{3}-\mu\ln\bigg(\frac{r}{2M}\bigg)
\end{equation} for which the $\beta$ term in the equation (\ref{exp})
reads
\begin{equation}\label{betar}e^{2\epsilon\beta(r)}\approx1+\frac{2m(r)}{r}+\frac{\Lambda}{3}r^2-\mu\ln\bigg(\frac{r}{2M}\bigg).\end{equation}
where
\begin{equation}\label{mass}m(r)=M\bigg(\frac{r}{2M}\bigg)^\delta\end{equation}
 is mass function of the black hole with
 \begin{equation}\label{delta}\delta=2-s_0=\frac{8\epsilon(5+\sigma\sqrt{1+\sigma})}{1-4\epsilon[9+4\sigma-(2-\sqrt{\sigma})\sqrt{\sigma(1+\sigma)}]}.
 \end{equation}
  By looking at the metric solution (\ref{betar}) we can infer that our obtained black hole metric solution approaches to a modified Schwarzschild de Sitter like black hole asymptotically
  in weak field limits. One can see that the mass function (\ref{mass}) can be reduce to a
point mass $M$ which is applicable in the Schwarzschild de Sitter
black hole. To do so we must be choose $s_0=2$ in the above metric
which is not a good choice because it reaches to condition
$\epsilon=0.$ Thus one can infer that the logarithmic corrections
on the metric potential and mass function are unavoidable two
results generated from gravitational effects of an accelerated
observer with four velocity (\ref{Nt}). In this sense one can
obtain radius of event horizon $r_e$
 for the above  black hole metric
solution by solving the equation $e^{-2\epsilon \beta(r_e)}=0$ as
follows.
\begin{equation}\end{equation}
\begin{equation}e^{-2\epsilon\beta(r_e)}\approx1-\bigg(\frac{2M}{r_e}\bigg)^{s_0-1}-\frac{\Lambda}{3}r_e^2+\mu\ln\bigg(\frac{r_e}{2M}\bigg)=0.
\end{equation}
We plot all possible solutions of the above horizon equation
versus the $\sigma$ parameter of the accelerating observer in
figures 1 and 2 for positive and negative numeric values of the
dimensionless cosmological parameter respectively in which
\begin{equation}x_e=\frac{r_e}{2M},~~~~~\lambda=\frac{4M^2\Lambda}{3}.\end{equation}
These diagrams show that relative distance of two horizons of this
black hole is depended to numeric value of vector field or
acceleration parameter of preferred reference frame and this
distance is decreased by increasing the acceleration parameter. In
fact there is particular numeric value for this parameter in which
both of black hole horizon and cosmological horizon reach to each
other.  After some talk about the initial constants and by
redefining  them with the physical parameters for $\beta(r) $ we
set $C_{1,2}=0$ and then we substitute (\ref{epsbet}) into
(\ref{C}) to obtain $\alpha(r)$ such that
\begin{equation}\label{cond}2\epsilon\alpha(r)=-4\epsilon\beta(r)=-2\bigg(\frac{2M}{r}\bigg)^{s_0-1}-\frac{2\Lambda r^2}{3}
+2\mu\ln\bigg(\frac{r}{2M}\bigg)\end{equation} and so $\alpha(r)$
term in the equation (\ref{exp}) become
\begin{equation}e^{2\epsilon\alpha(r)}\approx1+2\epsilon\alpha(r)=1-2\bigg(\frac{2M}{r}\bigg)^{s_0-1}-\frac{2\Lambda}{3}r^2+2
\mu\ln\bigg(\frac{r}{2M}\bigg).\end{equation} Substituting
(\ref{betac}) and (\ref{cond}) and condition
\begin{equation}\label{cond22}a+\epsilon p=0\end{equation} into the equation
(\ref{alpha}) we obtain
\begin{equation}r^2\psi^{\prime\prime}+2r\psi^\prime+4\psi=\end{equation}$$\frac{\gamma_1\Lambda r^2}{
3\epsilon}+\frac{\gamma_2}{2\epsilon
}\bigg(\frac{2M}{r}\bigg)^{s_0-1}-\frac{3\mu}{\epsilon}\ln\bigg(\frac{r}{2M}
\bigg)+\frac{\gamma_3\mu}{2\epsilon}-\frac{2(1+2\sigma)}{\epsilon}+8\sigma.
$$ The condition (\ref{cond22}) is important because it makes
possible an analytic solution for the above differential equation
as follows.
\begin{equation}\label{psir}\psi(r)=\sqrt{\frac{2M}{r}}\bigg\{C_5\sin\bigg(\frac{\sqrt{15}}{2}\ln\bigg(\frac{r}{2M}\bigg)\bigg)+C_6
\cos\bigg(\frac{\sqrt{15}}{2}\ln\bigg(\frac{r}{2M}\bigg)\bigg)\bigg\}\end{equation}
$$-\frac{3\mu}{4\epsilon}\ln\bigg(\frac{r}{2M}
\bigg)+\frac{\gamma_1\Lambda
r^2}{30\epsilon}+\frac{\gamma_2}{2\epsilon(s_0^2-3s_0+6)
}\bigg(\frac{2M}{r}\bigg)^{s_0-1}$$$$+\frac{\mu(3+2\gamma_3)+16\sigma(2\epsilon-1)-8}{16\epsilon}$$
where $C_{5,6}$ are integral constants and
\begin{equation}\label{gamma1}\gamma_1=2A_1+2A_3+3-A_2-A_4\end{equation}
\begin{equation}\gamma_2=6+(s_0-1)[A_4-2A_3+s_0(2A_1-A_2)]\end{equation}
\begin{equation}\gamma_3=2A_1-A_2-2A_3+A_4.\end{equation}
Looking at the definition of the Brans Dicke field (\ref{sigma})
and series expansion (\ref{exp}) and substituting the $\psi$
solution (\ref{psir}) we obtain
\begin{equation}G\phi(r)\approx\frac{1}{2}+\sigma(2\epsilon-1)+\frac{\mu(3+2\gamma_3)}{16}
-\frac{3\mu}{4}\ln\bigg(\frac{r}{2M}\bigg)\end{equation}$$+\frac{\gamma_2}{2(s_0^2-3s_0+6)}
\bigg(\frac{2M}{r}\bigg)^{s_0-1}+\frac{\gamma_1\Lambda
r^2}{30}+$$$$ \sqrt{\frac{2M}{r}}\bigg\{\epsilon
C_5\sin\bigg(\frac{\sqrt{15}}{2}\ln\bigg(\frac{r}{2M}\bigg)\bigg)+\epsilon
C_6
\cos\bigg(\frac{\sqrt{15}}{2}\ln\bigg(\frac{r}{2M}\bigg)\bigg)\bigg\}.
$$
Physically we know that $\phi(r)$ should be positive raising
function versus $r$ because inverse of the Brans Dicke scalar
field is the Newton`s gravity coupling parameter and in according
to the Mach`s principle $G(r)=Ge^{-\epsilon\psi(r)}$ should be
decreasing function \cite{BD}. In this sense for $\Lambda>0$ we
should choose $\gamma_1>0.$ Substituting the coefficients
(\ref{A1}), (\ref{A2}), (\ref{A3}) and (\ref{A4}) into
(\ref{gamma1}) we obtain
\begin{equation}\label{gamm}\gamma_1=27+18\sigma+\epsilon\{(16\sqrt{1+\sigma}-192)\sigma^2-4\sqrt{1+\sigma}(3+4\sqrt{\sigma})\sigma^\frac{3}{2}
\end{equation}$$
+(16\sqrt{1+\sigma}-32\sqrt{\sigma(1+\sigma)}-612)\sigma+\sqrt{\sigma(1+\sigma)}(64+64\sigma-16\sqrt{\sigma})$$$$-420+32\sqrt{\sigma(1+\sigma)}$$
and
\begin{equation}\label{mu}\mu(\sigma)=\frac{2\epsilon\sigma}{(1+\sigma)(1-\delta)[1-4\epsilon(9+4\sigma-\sqrt{\sigma(1+\sigma)(2-\sqrt{\sigma})})]}
.\end{equation} One can check for $\gamma_1>0$ and $\sigma>0$ the
equation (\ref{gamm}) reads to condition $\epsilon<\frac{1}{15}.$
Thus we will use ansatz
\begin{equation}\label{epsilon}\epsilon=\frac{1}{50}\end{equation}in what follows. Regarding (\ref{epsilon}) we plot diagrams
for $\gamma_1,\delta,\mu$ and $\omega$ versus $\sigma>0$ in figure
3. They show different possible metric solutions in presence of
the vector field effects. They show convergent mass function for
$\delta<0$ and divergent mass function for $\delta>0.$ However
diagram in figure 3-a shows to have $\gamma_1>0$ we must be choose
$0<\sigma<6.$ Regarding these boundary conditions on the
parameters of the metric solution we investigate gravitational
lensing of this obtained black hole metric via approaches of wave
optics and geometric optics in the next section.
\section{Theory of Fresnel-Kirchhoff diffraction}
We should remember that the Fresnel Kirchhoff diffraction is
applicable when source is far from the observer and so fringe of
light waves are formed on the telescope eyepiece .Hence this
diffraction will be applicable in study image formation from
gravitational lensing in which the stars (sources) and the black
holes (the lens) are far from us. The first we review the general
formalism of gravitational lensing via wave optics approximation.
Ignoring the polarization property the electromagnetic waves can
considered just as scalar waves which propagate on a curved space
time. For simplicity we choose here a massless scalar wave which
is defined by the following equation.
\begin{equation}\label{2}\partial_{\mu}(\sqrt{-g}g^{\mu\nu}\partial_{\nu}\Phi)=0\end{equation}in which $g_{\mu\nu}$
is background metric field which has role of a gravitational lens
and $g$ is absolute value of determinant of it.
 In weak gravitational field limits
we consider the background metric of a static object (the lens)
has the following form in units $c=G=1$.
\begin{equation}\label{line}
 ds^2\approx-(1+2U(\tilde{r}))dt^2+(1-2U(\tilde{r}))(d\tilde{r}^2+\tilde{r}^2(d\theta^2+\sin^2\theta d\varphi^2))
\end{equation} where
 $|U(\tilde{r})|<<1$ and for our model is given by (\ref{epsbet}) as follows.
 \begin{equation}\label{U} 2U(\tilde{r})=-\bigg(\frac{2M}{\tilde{r}}\bigg)^{1+\delta}-
 \frac{\Lambda \tilde{r}^2}{3}+\mu\ln\bigg(\frac{\tilde{r}}{2M}\bigg)
  \end{equation} in which $\tilde{r}$ is the radial
  coordinate evaluated from center of the black hole (the lens plane in figure 4).
 The line element (\ref{line}) is in fact the Minkowskian flat metric plus small disturbance due to the gravitational
potential $U(\tilde{r}).$ To study diffraction of light of a star
located at the source plane (see figure 4) via the black hole we
must first use a local spherical coordinates system which its
center is in the source position and its polar axis is pointing
toward the lens (black hole) and then we act to solve wave
equation  (\ref{2}). In this sense $U(\tilde{r})$ should be
changed by
$U(|\vec{r}-\vec{D}_{LS}|)=U(r,\overrightarrow{\theta})$ in which
$\vec{D}_{LS}$ is vector distance between source and lens in
figure 4,
$\overrightarrow{\theta}=\theta(\cos\varphi,\sin\varphi)$ is a two
dimensional vector on a flat lens plane and radial coordinate $r$
should be evaluated from the source position instead of the lens
position (see figure 4) so that
\begin{equation}\label{tildar}\tilde{r}=|\overrightarrow{r}-\overrightarrow{D}_{LS}|=\sqrt{r^2-2rD_{LS}
 \cos\theta+D_{LS}^2}.\end{equation}
Substituting
\begin{equation}\label{3}\phi(t,\vec{r})=\tilde{\phi}(\vec{r})e^{-i\varpi t}\end{equation}
into the wave equation (\ref{2}) and eliminating higher order
terms in $U$ one can obtain equation of the amplitude
$\tilde{\phi}(\vec{r})$ as
\begin{equation}\label{4}(\nabla^2+\varpi^2)\tilde{\phi}=4\varpi^2U(\vec{r})\tilde{\phi}\end{equation}
where $\nabla^2$ is the flat space Laplacian. By using the
spherical polar coordinates system the amplitude equation
(\ref{4}) reads
\begin{equation}\label{Lap}\frac{1}{r}\frac{\partial^2}{\partial r^2}(r\tilde{\phi})+\frac{1}{r^2\sin\theta}\frac{\partial}{\partial\theta}
\bigg(\sin\theta\frac{\partial\tilde{\phi}
}{\partial\theta}\bigg)+\frac{1}{r^2\sin^2\theta}\frac{\partial^2\tilde{\phi}}{\partial\varphi^2}+\varpi^2\tilde{\phi}(\vec{r})=4\varpi^2
U(\vec{r})\tilde{\phi}(\vec{r}).\end{equation} We assume the wave
scattering occurs in a small spatial region around the lens and
outside of this region the wave propagates in a flat space. In
other words one can see that without the lensing object for which
$U=0$ the above equation has a simple solution as
\begin{equation}\label{6}\tilde{\phi}_0=\frac{\Omega e^{i\varpi r}}{r}\end{equation} in which $\Omega$ is a constant and $r=|\vec{r}|$.
Hence it is useful we act to define the amplification factor of
the wave amplitude due to lensing as follows.
\begin{equation}\label{7}F(\vec{r})=\frac{\tilde{\phi}(\vec{r})}{\tilde{\phi}_0(\vec{r})}.\end{equation}
We set the observer is located at
$\vec{r}_o=(r_o,\theta_o,\varphi_o)$ with $\theta_o<<1 (radians)$
on the observer plane (see figure 4). In this case the waves which
is detected by the observer should be confined in the small
$\theta<<1$ regions with $0\leq\varphi\leq2\pi$ for which the
approximation $\sin\theta\approx\theta$ can be used in the
equation (\ref{Lap}) . Substituting (\ref{6}) and (\ref{7}) and
the approximation $\sin\theta\approx \theta,$ the equation
(\ref{Lap}) reduces to the following form.
\begin{equation}\label{Lap2}\frac{\partial^2F}{\partial r^2}+2i\varpi\frac{\partial
F}{\partial r}+\frac{\nabla^2_{\theta }F}{r^2}=4\varpi^2 U (r,\theta)F
\end{equation}
where we defined
\begin{equation}\label{Lap3}\nabla^2_{\theta}=\frac{\partial^2}{\partial\theta^2}+\frac{1}{\theta}\frac{\partial}{\partial\theta}+
\frac{1}{\theta^2}\frac{\partial^2}{\partial\varphi^2}.\end{equation}
 According to the work presented in Refs. \cite{RT,YN} we choose geometry of the gravitational lens to be made by three parallel planes,
 namely the source plane, the lens plane and the observer plane (see figure 4). The angular diameter distances along the normal from the observer
 plane to the source and the lens planes are labeled with $D_{SO}$ and  $D_{LO}$, respectively. Distance between the source and the lens planes is
 labeled as $D_{LS}.$ The emitted waves by a point source located in the source plane, travel freely to the lens plane, and they are lensed by
 a gravitational potential $U$ where they are assumed to be localized in the thin (width$<<$ focal length) lens plane , before  reaching the telescope
  at the observer plane. We assume  $\overrightarrow{\xi}$ is the coordinates at the lens plane,  $\overrightarrow{\eta}$ is the coordinates at
  the source plane and $\vec{\Delta}$ is
 coordinates at the observer plane in figure 4, which can be redefined  by the following dimensionless coordinates respectively.
 \begin{equation}\label{diff} \overrightarrow{x}=\frac{\overrightarrow{\xi}}{\xi_0}=\frac{\vec{\theta}}{\theta_E},
 ~~~\overrightarrow{y}=\frac{D_{LO}}{D_{SO}}
 \frac{\overrightarrow{\eta}}{\xi_0}=\frac{\vec{\beta}}{\theta_E},~
 ~~~~
 \vec{z}=\bigg(1-\frac{D_{LO}}{D_{SO}}\bigg)\frac{\vec{\Delta}}{\xi_0}\end{equation}
 where $\xi_0=D_{LO}\theta _E$ is radius of the Einstein rings and $\theta_E$ is corresponding angular
  radius which for a point lens with mass $M_L$ is given by the following equation.
\begin{equation}\label{thetae}\theta_E=2\sqrt{\frac{M_L}{D_{LO}}\frac{D_{LS}}{D_{SO}}}.\end{equation}
 Applying (\ref{diff}) and (\ref{thetae})
one can show that the equation (\ref{Lap2}) has an integral
solution (see \cite{YN} and \cite{TT}) which at the observer plane
is given by
\begin{equation}\label{fy}F(\vec{y}, \vec{z})=\frac{w}{2\pi i}\int dx^2e^{iwT(x,y, z)}.\end{equation}
This integral solution is called as Fresnel-Kirchhoff diffraction
formula. The dimensionless time delay function $T(x,y,z)$ and
dimensionless frequency $w$ are given respectively by the
following relations.
\begin{equation}\label{Txyz}
T(x,y,z)=\frac{1}{2}|\overrightarrow{x}
-\overrightarrow{y}-\vec{z}|^2-\Psi(\overrightarrow{x}),~~~~w=\frac{D_{SO}\xi_0^2}{D_{{LS}}D_{LO}}\varpi=4M_L\varpi\end{equation}
in which 2 dimensional lens potential is defined by the following
equation.
\begin{equation}\label{lpot}\Psi(\overrightarrow{x})=\frac{1}{2M_L}\int_{\ell}^{D_{SO}}dr U(\overrightarrow{x},r)\end{equation}
where $\ell$ is scale of source. In fact it is obtained by
projecting the three dimensional metric potential
$U(r,\vec{\theta})$ on the lens plane. This lens potential
satisfies two important properties as follows:  According the
following equation its gradient become deflection angle.
\begin{equation}\label{15}\overrightarrow{\nabla_x}\Psi(\overrightarrow{x})=\overrightarrow{\alpha}(\overrightarrow{x})\end{equation} and
 its two dimensional
Laplacian gives twice of the convergence as
\begin{equation}\label{16} \nabla^2_{x}\Psi(\overrightarrow{x})=2\kappa(\overrightarrow{x})\end{equation}
where
\begin{equation}\label{17} \kappa(\overrightarrow{x})=\frac{\Sigma(\overrightarrow{x})}{\Sigma_{cr}}\end{equation}
is in fact the dimensionless surface gravity
$\Sigma(\overrightarrow{x})$ of the lens evaluated on the lens
plane and
\begin{equation}\label{18}\Sigma_{cr}=\frac{D_{SO}}{4\pi D_{LO}D_{LS}}\end{equation} is called the critical surface gravity. The time delay
$T(x,y,z)$ can be obtained by applying the path integral formalism
on the possible pathes of the moving light rays by regarding the
eikonal approximation in the equation (\ref{Lap2}). In the eikonal
approximation one can neglect second order differentiation with
the first one in the equation (\ref{Lap2}) because of the
assumption $\varpi/|\partial \ln F/\partial r|\sim$(Scale on which
$F$ varies)/(wavelength)$>>1$ and so the equation (\ref{Lap2})
looks like the Schrodinger equation \cite{TT}.
 In this latter case the $r$ component behaves same as the time
evolution parameter for the wavefronts with particle mass
$\varpi$. In the geometrical optics limit $\varpi>>1$ the
diffraction integral (\ref{fy}) reaches to the stationary points
of the phase function which they are obtained by the Fermat`s
principle as follows.
\begin{equation}\label{19}\overrightarrow{\nabla_x}T(x,y,z)=0=\overrightarrow{x}-\overrightarrow{y}-
\vec{z}-\overrightarrow{\alpha}(\overrightarrow{x})\end{equation}
in which the source position $\overrightarrow{y}$ is fixed.  This
is lens equation in geometrical approximation for gravitational
lensing and determines the location of the images
$\overrightarrow{x}$ for a given source position
$\overrightarrow{y}.$  At the observer plane the unlensed wave
amplitude (\ref{6}) is $\phi_o(\vec{\eta},
\vec{\Delta})=(\Omega/r_s)\exp(i\varpi r_s)$ where
$r_s=\sqrt{D_{SO}^2+|\vec{\eta}- \vec{\Delta}|^2}.$ At the
Fresnel-Kirchhoff diffraction limit for which
$D_{SO}>>|\vec{\eta}- \vec{\Delta}|$  we can write
$\phi_o(\vec{\eta},
\vec{\Delta})\approx(\Omega/D_{SO})\exp[i\varpi(D_{SO}+|\vec{\eta}-
\vec{\Delta}|^2/2D_{SO})]$ and then we substitute the
dimensionless coordinates (\ref{diff}) to obtain \cite{RT,YN}
\begin{equation}\label{20}\phi_0(\vec{y},\vec{z})=\frac{\Omega e^{iwt(\vec{y},\vec{z})}}{D_{SO}}.\end{equation} In the above relation
 \begin{equation}\label{21}t(\vec{y},\vec{z})=\frac{1}{2}\sqrt{\frac{p}{1-p}}\bigg|\sqrt{\frac{p}{1-p}}\vec{y}-\sqrt{\frac{1-p}{p}}\vec{z}\bigg|^2+
 \bigg(\frac{1-p}{2q}\bigg)\end{equation}
and
\begin{equation}\label{pq}p=\frac{D_{LS}}{D_{SO}},~~~q=\frac{2M_L}{D_{SO}},~~~\end{equation} and so the lensed
wave amplitude will be
\begin{equation}\label{22}\phi_L(\vec{y},\vec{z})=\phi_0(\vec{y},\vec{z})F(\vec{y},\vec{z}).\end{equation}
This is an entrance  wave which enters in front of the telescope
convex lens (see figure 4-a) and so part of it which passes
 through the lens makes the transmitted wave at the observer plane (the telescope) which is given by the following equation.
 \begin{equation}\label{23}\phi_T(\vec{y},\vec{z'})=\tau(\vec{z^\prime})e^{-i\varepsilon
  w|\vec{z'}|^2}
 \phi_L(\vec{y},\vec{z'}),~~~\varepsilon=\frac{D_{LO}D_{SO}}{2fD_{LS}}.\end{equation}  Here $z^\prime$ and $f$ are position and focal length
  of the telescope convex lens respectively.  The function of aperture of the convex lens $\tau(\vec{z^\prime})$ is defined by
  $\tau(\vec{z})=1$ for $0\leq|\vec{z}|\leq R$
  and $\tau(\vec{z})=0$ for $|\vec{z}|>R$ in which $R$ is lens radius of the telescope. We apply the lens equation of a convex thin lens as
  \begin{equation}\label{24}\frac{1}{D_{SO}}+\frac{1}{z_o}=\frac{1}{f}\end{equation}
 to obtain amplitude of the magnified wave which after passing from the
 lens causes to form the image at the position $zo$ on the observer plane. This is done by the integral equation
\begin{equation}\label{25}\phi_o(\vec{y}, \vec{z})=\int_{|
\vec{z^\prime}|\leq
R}dz^{\prime2}\phi_T(\vec{y},\vec{z'})e^{i(\varepsilon
wf/z_o)|\vec{z}-\vec{z^\prime}|^2}.\end{equation} By substituting
(\ref{20}), (\ref{22}) and (\ref{23}) into the equation (\ref{25})
we obtain
\begin{equation}\label{phio} \phi_o(\vec{y}, \vec{z})=\frac{\Omega\exp[iwg(\vec{y},\vec{z})]}{D_{SO}}\int_{|\vec{z^\prime}|\leq
R}dz^{\prime2}F(\vec{y},\vec{z'}) e^{-iw\vec{k}
\cdot\vec{z^{\prime}}}
\end{equation}  where \begin{equation}\label{gg}g(\vec{y},\vec{z})=\bigg(\frac{p}{1-p}\bigg)^\frac{3}{2}\frac{y^2}{2}+\frac{D_{LO}}{2pz_o}z^2
+\frac{1-p}{2q}
\end{equation}
and
\begin{equation}\label{28}\vec{k}=\sqrt{\frac{p}{1-p}}\vec{y}+\frac{D_{LO}}{pz_o}\vec{z}\end{equation}
and in the exponent we omitted $|\vec{z^\prime}|^2$ with respect
to the linear orders . Because evaluate of the integral is in a
small region $|z^\prime|\leq R$ in which dimensionless radius of
the telescope convex lens is small $R<<1.$ In fact $R$ is the
convex lens radius divided by $D_{SO}.$ On the other side we know
that distance between the source and the observer is very large
versus radius of the convex lens in the Fraunhofer diffraction.
However one can look at the equation (\ref{phio}) to infer that
$\phi_o(\vec{y}, \vec{z})$ is in fact the Fourier transform of the
amplification factor $F$ which  produces the interference fringe
pattern. Furthermore the lens
 equation (\ref{24}) shows that for large distances $D_{SO}>>1$ we will have \begin{equation}\label{29}z_o\approx f.\end{equation}
  By substituting the above approximation into (\ref{gg}) and (\ref{28}) and by applying (\ref{Txyz}) and (\ref{fy}), the equation (\ref{phio})
   reaches to the following form.
\begin{equation}\label{phioj}\phi_o(\vec{y},\vec{z})\approx
\frac{\Omega w\exp[iwg(\vec{y},\vec{z})]}{2\pi iD_{SO}}\int dx^2
\exp[iw(|\vec{x}-\vec{y}|^2/2-\psi(\vec{x}))]\times\end{equation}$$\int_0^Rz^\prime
dz^\prime\int_0^{2\pi}d\chi\exp(-iw \rho z^{\prime}\cos\chi)$$
where $\chi$ is angle between $\vec{\rho}$ and $\vec{z}^\prime$
and
\begin{equation}\vec{\rho}=\vec{x}+\bigg(\sqrt{\frac{p}{1-p}}-1\bigg)\vec{y}+\frac{D_{LO}}{pf}\vec{z}\end{equation}
 and we omitted again $z^{\prime2}$ with respect to its first order term because of above mentioned reasons.
  By applying integral form of the zero order Bessel function
  \cite{GA}
\begin{equation}\label{31} J_0(s)=\frac{1}{2\pi}\int_0^{2\pi} d\chi e^{-is\cos\chi}
\end{equation} one can calculate the angular  $\chi$ part in the integral equation (\ref{phioj}) and obtain the zero order Bessel
 function $J_0(w\rho z^\prime)$ and then by using the identity
\begin{equation}sJ_0(s)=\frac{d(sJ_1(s))}{ds}\end{equation} at last the
equation (\ref{phioj}) reaches to the following form.
\begin{equation}\label{phi}\phi_o(\vec{y},\vec{z})=\frac{\Omega R\exp[iwg(y,z)]}{iD_{SO}}\int dx^2\frac{J_1(wR\rho)
}{\rho}e^{iw(|\vec{x}-\vec{y}|^2/2-\Psi(x))}.\end{equation}
 At the geometrical optics limit where $w>>1$ the amplification factor (\ref{fy}) approaches to the  WKB approximated form as
\begin{equation} \label{32}F(x_s,z)\sim \frac{w}{2\pi i}e^{iw(|\vec{x}_s-\vec{y}-\vec{z}|^2/2-\psi(\vec{x_s}))}\end{equation}
where positions of the stationary images $\vec{x}_s$ are obtained
from the lens equation (\ref{19}) and so the image amplification
factor (\ref{phi}) reduces to the following form.
\begin{equation}\label{33}\phi_o(\vec{z},\vec{x}_s)\sim\exp\{i\frac{w}{2}[2g(\vec{y},\vec{z})+|\vec{x}_s-\vec{y}-\vec{z}|^2-2\psi(\vec{x_s})]\}
\frac{J_1(wR\rho_s)}{wR\rho_s}\end{equation} in which we should
use (\ref{19}) to obtain images position $x_s$ versus the fixed
source position $y$ and replace into the above equation. Also the
above Bessel function reduces to a two-dimensional Dirac delta
function in geometric optics limit $wR>>1$ such that we can use
\begin{equation}\label{34}\lim_{wR\to\infty}
\phi_o(\vec{y},\vec{z})\sim\delta^2(wR\rho_s).
\end{equation}
This amplitude of the wave gives us position of the images at the
observer plane in the geometrical optics limit $\rho_s=0$ which
can be rewritten as follows.
\begin{equation}\label{35}\vec{z}_s=\frac{pf}{D_{LO}}\bigg[\bigg(1-\sqrt{\frac{p}{1-p}}\bigg)\vec{y}-\vec{x}_s\bigg].\end{equation}
 If the lens equation (\ref{19}) has multiple stationary image positions $\vec{x}_s^i$ with
 $i=1,2,...$ then the amplified waves on the observer plane for these stationary images reaches \cite{YN} to the following form.
\begin{equation}\label{36}\phi_o(\vec{z})=\sum_{i=1}\Upsilon_i\phi_o(\vec{z}, \vec{x}_s^i)\end{equation} where amplitude $\Upsilon_i$ should
be set as magnitude of image magnification factor in the geometric
optics limit as follows.
\begin{equation}\label{Ups}\Upsilon_i=\big(\frac{y}{x}\frac{dy}{dx}\big)^{-1}_{|_{x=x_s^i}}.\end{equation}
In the next section we use metric potential of our model
(\ref{line1})  to calculate corresponding lens potential and then
to produce the images via approaches of geometric optics and wave
optics.
\section{Lens potential}
  By substituting (\ref{tildar})
and  (\ref{U}) into the integral equation of the lens potential
(\ref{lpot}) we obtain
 \begin{equation}\label{int}\Psi(\theta)=-\frac{1}{2q}\int_{0}^1dx\{\bigg(\frac{q}{\sqrt{x^2-2xp\cos\theta+p^2}}\bigg)^{1+\delta}+\end{equation}$$
 h(x^2-2xp\cos\theta+p^2)-\mu\ln(\sqrt{x^2-2xp\cos\theta+p^2}/q)\}$$
where we used (\ref{pq}) and the following definitions
\begin{equation}~~~~h=\frac{\Lambda D_{SO}^2}{3},~~~x=\frac{r}{D_{SO}}\end{equation}
and assumed $\ell=0$ because of small scale of the cosmic source
from point of view of earth observer. However one can check that
the above integral has an analytic solution just for integer
numeric values  for $\delta$ parameter and before to calculate it
we must be substitute numeric value for $\delta.$ All possible
numeric values for $\delta$ parameter are given in figure (3-b) .
We will need asymptotic behavior of the lens potential at
$\sin\theta\approx\theta$ which we used previously in the
Laplacian (\ref{Lap2}). In this sense one can check that the lens
potential is singular namely at limits $\theta\to0$ it diverges to
infinity just for $\delta>0$ but not for $\delta<0.$ First term in
Taylor series of lens potential of our model begins by
$\frac{1}{\theta^\delta}$ in which $\delta>0.$ We also claim that
$\delta<0$ gives not physical metric solutions because the
function of black hole mass (\ref{mass}) should be an increasing
function by raising $r.$ In these sense we calculate Taylor series
expansion of the integral equation (\ref{int}) for ansatz
$\delta=0,1,2,3$ respectively as follows.
\begin{equation}\label{psi0}\Psi^{\delta=0}(\theta)=\Psi_0^{0}+\ln\theta+\Psi_1^0\theta+O(\theta^2
)\end{equation} where
\begin{equation}\Psi_0^0=-\frac{1}{2q}\bigg[\mu+\frac{h}{12}+\frac{(1-\mu)\pi i}{4}+\ln (q2^{1+2q})\bigg],~~~\Psi_1^0=\frac{\mu\pi}{4q}\end{equation}
\begin{equation}\label{psi1} \Psi^{\delta=1}(\theta)\approx\frac{\Psi^1_{-1}}{\theta}+\Psi^1_0+\Psi^1_1\theta+O(\theta^2)\end{equation}
where
\begin{equation}\Psi^1_{-1}=-\frac{\pi q}{2p},~~~\Psi^1_1=\frac{\pi(6\mu p-q^2)}{12q}\end{equation} and
\begin{equation}\Psi^1_0=\frac{1}{2q}\bigg\{\frac{q^2(p-2)}{p(p-1)}-\frac{h}{3}+hp-hp^2+\mu\ln\bigg(\frac{p^p}{(1-p)^{p-1/2}}\bigg)\bigg\}.\end{equation}
\begin{equation}\label{psi2}\Psi^{\delta=2}(\theta)=\frac{\Psi_{-2}^2}{\theta^2}+\Psi^2_0+\Psi^2_1\theta+O(\theta^2)\end{equation}
where
\begin{equation}\Psi_{-2}^2=-\frac{q^2(1+p)}{2p^2},~~~\Psi_1^{2}=\frac{\mu\pi p}{2q}\end{equation} and
\begin{equation}\Psi_0^2=\frac{1}{2q}\bigg\{\frac{q^3}{2(1-p)^2}-\frac{q^3(2+p)}{64p^2(1-p)}-\frac{h}{3}+hp-hp^2-\mu
+\end{equation}$$\mu\ln\bigg(\frac{p^p(1-p)^{1-p}}{q}\bigg)+\frac{q^2}{12p^2}
\bigg\}.$$
\begin{equation}\label{psi3}\Psi^{\delta=3}(\theta)=\frac{\Psi^3_{-3}}{\theta^3}+\frac{\Psi_{-2}^3}{\theta^2}+\frac{\Psi_{-1}^3}{\theta}+\Psi_0^3+\Psi_1^3\theta+O(\theta^2
)\end{equation} where we defined
\begin{equation}\Psi_{-3}^3=-\frac{\pi q^3}{4p^3},~~~\Psi_{-2}^2=-\frac{q^3(1-2p)}{4p^3(1-p)^2},~~~\Psi_{-1}^3=-\frac{\Psi_{-3}^3}{2}\end{equation}
and
\begin{equation}\Psi_0^{3}=\frac{1}{2q}\bigg\{\frac{q^4}{3p^3}+\frac{q^4}{3(1-p)^3}-\frac{h}{3}+hp-hp^2-\mu+\mu\ln\bigg(\frac{p^p(1-p)^{1-p}}{q
}\bigg)\bigg\}\end{equation}\begin{equation}\Psi_{1}^3=\frac{\pi(240\mu
p^4-17q^4)}{480qp^3}.\end{equation}
 By looking at the above lens
potentials we infer that $\delta=0$ with logarithmic singularity
at $\theta\to0$ corresponds to a point lens while
$\delta=1,2,3,\cdots$ correspond with non point lens which their
spatial distributions of mass are given by (\ref{mass}). Looking
at these lens potentials one can infer that the vector field
effects with $\mu(\sigma)\neq0$ (see eq. (\ref{mu})) remove
logarithmic singularity but creates a power law singularity
$\theta^{-\delta}$ for $\delta>0$ for the lens potential. It is
obvious this will affect the position of the images.  We are now
in position to calculate images magnification factor (\ref{fy})
for the above lens potentials as follows.
\subsection{Images magnification factor}
To calculate the magnification factor (\ref{fy}) of the lensed
waves, we choose a two dimensional polar coordinates on the lens
plane for which the surface differential is $dx^2=x dx d\zeta$
where $\zeta$ is polar angle between directions of the vectors $
\vec{x}$ and $\vec{y}+\vec{z}$ on the lens plane so that
$|\vec{x}-(\vec{y}+\vec{z})|^2=x^2+|\vec{y}+\vec{z}|^2-2x|\vec{y}+\vec{z}|\cos\zeta.$
Applying these relations the equation (\ref{fy}) can be rewritten
as follow.
\begin{equation}\label{65}F(y,z)=-iw
e^{(iw/2)|\vec{y}+\vec{z}|^2}\int_0^\infty x dx
J_0(wx|\vec{y}+\vec{z}|)e^{iw[\frac{x^2}{2}-\Psi(x)]}\end{equation}
in which  we substituted the following identity for the zero order
Bessel function.
\begin{equation}\label{66}J_0(wx|\vec{y}+\vec{z}|)=\frac{1}{2\pi}\int_0^{2\pi}d\zeta e^{-iwx|\vec{y}+\vec{z}|\cos\zeta}.\end{equation}
Substituting the series solution \cite{GA}
\begin{equation}\label{67}
J_0(s)=\sum_{k=0}^{\infty}\frac{(-1)^{k}}{(k!)^2}\bigg(\frac{s}{2}\bigg)^{2k}\end{equation}
the integral equation (\ref{65}) reads
\begin{equation}\label{Fyz}F(y,z)=-iw
e^{(iw/2)|\vec{y}+\vec{z}|^2}\sum_{k=0}^{\infty}\frac{(-1)^k}{(k!)^2}
C_k\bigg(\frac{w|\vec{y}+\vec{z}| }{2}\bigg)^{2k}\end{equation}
where we defined
\begin{equation}\label{F0}C_k=\int_0^\infty
x^{1+2k}e^{iw[x^2/2-\Psi(x)]}dx.
\end{equation}
We see that numeric values of the coefficients $C_k$ are dependent
to form of the lens potential. By substituting $\theta=\theta_E x$
into the lens potentials (\ref{psi0}), (\ref{psi1}) and
(\ref{psi2}) and (\ref{psi3}) and by keeping just highest order
singular term in these potentials to calculate (\ref{F0}) and by
using the integral form of the gamma function  $n!=\int_0^\infty
t^ne^{-t}dt$ \cite{GA} at last we obtain
\begin{equation}\label{Ck0}C^{\delta=0}_k\approx e^{-iw(\Psi^0_0+2\ln\theta_E)}2^{1+k
-iw/2}\bigg(\frac{i}{w}\bigg)^{1+k-iw/2}\bigg(k-\frac{iw}{2}\bigg)!\end{equation}
for point source $\delta=0$ and for non point sources
$\delta=1,2,3$ we have
\begin{equation}\label{Ck1}C_k^{\delta}\approx e^{-iw\Psi^\delta_0}\int_0^\infty dx x^{1+2k}e^{iw\Pi(x)}\end{equation}
where
\begin{equation}\Pi(x)=\frac{x^2}{2}-\frac{\Psi_{-\delta}^\delta}{\theta_E^\delta}\frac{1}{x^\delta}
\end{equation} has Taylor series expansion around its minimum point
\begin{equation}x_m=\bigg(\frac{-\delta\Psi_{-\delta}^\delta}{\theta_E^\delta}\bigg)^\frac{1}{2+\delta},~~~\theta_E=\sqrt{\frac{2pq}{1-p}}\end{equation}
as follows.
\begin{equation}\label{Pi}\Pi(x)\approx\bigg(\frac{2+\delta}{2\delta}\bigg)x_m^2+\frac{(1+\delta)}{2}(x-x_m)^2+O(3).\end{equation} Substituting
(\ref{Pi}) into  (\ref{Ck1}) one can show
\begin{equation}\label{Cdelta}C^{\delta}_k\approx e^{iw[(2+\delta)x_m^2/2\delta-\Psi_0^\delta]}\int_0^\infty
dxx^{1+2k}e^{iw(1+\delta)(x-x_m)^2/2}.\end{equation} Because
$|x_m|<1$ then we can use Taylor series of the above integral
equation around $x_m$ and substitute integral form of the gamma
function to obtain a series form of the solution for the
coefficients (\ref{Cdelta}). This is done as follows.
\begin{equation}\label{cxm}C^{\delta}_k\approx e^{iw[(2+\delta)x_m^2/2\delta-\Psi_0^\delta]}\bigg(\frac{i}{w(1+\delta)}\bigg)^{1+k}2^k k!\times
\end{equation}$$\bigg\{1+x_m\frac{(k+\frac{1}{2})!}{k!}\bigg(\frac{2w(1+\delta)}{i}\bigg)^\frac{1}{2}+x_m^2[1-iw(1+\delta)(1+k)]+O(x_m^3)\bigg\}.$$
By substituting (\ref{Ck0}) into the image magnification factor
 (\ref{Fyz}) we will have
\begin{equation}\label{Fp0}F_{\delta=0}(y,z)=2e^{-iw[2\ln\theta_E+\Psi_0^0-|\vec{y}+
\vec{z}|^2/2]}\bigg(\frac{2i}{w}\bigg)^{-iw/2}\times\end{equation}$$\sum_{k=0}^{\infty}\frac{(k-iw/2)!}{(k!)^2}
\bigg(\frac{-iw|\vec{y}+\vec{z}|^2}{2}\bigg)^k.
$$
 In usual
astrophysical situations we know that \cite{TT}
\begin{equation} \label{71}w=4M_L\omega\sim10^5\times\bigg(\frac{M_L}{M_\odot}\bigg)^\frac{1}{2}(\nu/GHz)>>1 \end{equation}
 and so it is useful we study asymptotic behavior of the  amplification factor (\ref{Fp0}) for $w>>1$ as follows. Consider
\begin{equation}\label{75}\frac{\big(k-\frac{iw}{2}\big)!}{\big(-\frac{iw}{2}\big)!}=\bigg(-\frac{iw}{2}\bigg)\bigg(-\frac{iw}{2}+1\bigg)
\bigg(-\frac{iw}{2}+2\bigg)\cdots
\bigg(-\frac{iw}{2}+k\bigg)$$$$=\bigg(-\frac{iw}{2}\bigg)^k\bigg(1+\frac{2}{iw}\bigg)\bigg(1+\frac{4}{iw}\bigg)\cdots\bigg(1+\frac{2k}{iw}\bigg)
\end{equation} which for large frequency $w>>1$ one can omit the negligible
terms so that
$\big(1+\frac{2}{iw}\big)\big(1+\frac{4}{iw}\big)\cdots\big(1+\frac{2k}{iw}\big)$
and so we can write
\begin{equation}\label{iden1}\lim_{w\to\infty}\frac{\big(k-\frac{iw}{2}\big)!}{\big(-\frac{iw}{2}\big)!}\cong \bigg(-\frac{iw}{2}\bigg)^k.\end{equation}
On the other side for large frequencies $w>>1$ the following
identity is satisfied.
\begin{equation}\label{iden2}\lim_{w\to\infty}\bigg(\frac{-iw}{2}\bigg)!\approx\bigg(\frac{-iw}{2}\bigg)^{-\frac{iw}{2}}.\end{equation}
By substituting (\ref{iden1}) and (\ref{iden2}) into the equation
(\ref{Fp0}) we obtain
\begin{equation}\label{FPL}F_{\delta=0}(y,z,w>>1)\approx2e^{-iw[2\ln\theta_E+\Psi_0^0-|\vec{y}+
\vec{z}|^2/2]}J_0(w|\vec{y}+\vec{z}|)\end{equation}in which we
used series form for the zero order Bessel function namely
$$J_0(w|\vec{y}+\vec{z}|)=\sum_{k=0}^{\infty}\frac{(-1)^k}{(k!)^2}\bigg(\frac{w}{2}|\vec{y}+\vec{z}|\bigg)^{2k}.$$
By substituting (\ref{cxm}) into the image magnification factor
(\ref{Fyz}) and then by integrating as step by step for each order
of Taylor series at last we obtain the following solution for
$\delta=1,2,3,\cdots$.
\begin{equation}\label{F}F_{\delta=1,2,3,\cdots}(y,z)=F^{(0)}(y,z)+x_mF^{(1)}(y,z)+x_m^2F^{(2)}(y,z)+O(x_m^3)\end{equation}
where
\begin{equation}F^{(0)}(y,z)=\frac{1}{(1+\delta)}\exp\bigg\{iw\bigg[\frac{\delta|\vec{y}+\vec{z}|^2}{2(1+\delta)}
+\frac{(2+\delta)}{2\delta}x_m^2-\Psi_0^\delta
\bigg]\bigg\}\end{equation} for which we used the identity
$e^s=\sum_{n=0}^\infty s^n/n!,$
\begin{equation}F^{(1)}(y,z)=\sqrt{\frac{w}{1+\delta}}\times\end{equation}$$
\exp\bigg\{-\frac{i\pi}{4}+iw\bigg[\frac{1}{2}|\vec{y}+\vec{z}|^2+\frac{(2+\delta)}{2\delta}x_m^2-\Psi_0^\delta
\bigg]\bigg\}\sum_{k=0}^\infty\frac{(k+\frac{1}{2})!}{(k!)^2}\bigg(\frac{-iw|\vec{y}+\vec{z}|^2}{2(1+\delta)}\bigg)^k$$
and
\begin{equation}F^{(2)}(y,z)=F^{(0)}(y,z)-\end{equation}
$$iw\exp\bigg\{iw\bigg[\frac{1}{2}|\vec{y}+\vec{z}|^2+\frac{(2+\delta)}{2\delta}x_m^2-\Psi_0^\delta
\bigg]\bigg\}\sum_{k=0}^\infty\frac{(k+1)}{k!}\bigg(\frac{-iw|\vec{y}+\vec{z}|^2}{2(1+\delta)}\bigg)^k.$$
By applying the approximations
\begin{equation}\lim_{k\to\infty}\frac{(1+k)}{k!}=\frac{(1+k)!}{(k!)^2}\approx\frac{1}{k!},~~~\lim_{k\to\infty}\frac{(k+\frac{1}{2})!}{(k!)^2}\approx\frac{1}{k!}\end{equation}
one can show
\begin{equation}F^{(1)}(y,z)=\sqrt{2w(1+\delta)}\exp\bigg(\frac{-i\pi}{4}\bigg)F^{(0)},\end{equation}$$~~~F^{(2)}(y,z)=[1-iw(1+\delta)]F^{(0)}(y,z)$$
and so we can write
\begin{equation}\label{amdel}F_{\delta=1,2,3,\cdots}(y,z)=\frac{\{1+\sqrt{2w(1+\delta)}\exp(-i\pi/4)
x_m+[1-iw(1+\delta)]x_m^2\}}{(1+\delta)}\times\end{equation}$$\exp\bigg\{iw\bigg[\frac{\delta|\vec{y}+\vec{z}|^2}{2(1+\delta)}+\frac{(2+\delta)}{2\delta}x_m^2-\Psi_0^\delta
\bigg]\bigg\}$$ where we have
\begin{equation}x_m^{\delta=1}=\bigg[\frac{\pi \sqrt{q}}{2p}\bigg(\frac{1-p}{2p}\bigg)^\frac{1}{2}\bigg]^\frac{1}{2},~~~x_m^{\delta=2}=\bigg(\frac{q(1-p^2)}{2p^3}\bigg)^\frac{1}{3}
\end{equation}
$$x_m^{\delta=3}=\bigg[\frac{3\pi
q^{\frac{3}{2}}}{4p^3}\bigg(\frac{1-p}{2p}\bigg)^\frac{3}{2}\bigg]^\frac{1}{4},~~~\Psi_{-1}^1=-\frac{\pi
q}{2p},$$$$~~~\Psi_{-2}^{2}=-\frac{q^2(1+p)}{2p^2},~~~\Psi_{-3}^{3}=-\frac{\pi
q^3}{4p^3}.$$ At the astrophysical scales we know that $p<1$ and
$q<<1$ and so $x_m$ defined by the above relations will be have
small numeric values. Hence we use ansatz
\begin{equation}\label{ansatz}p=\frac{1}{2},~~~q=\frac{1}{10}\end{equation}
for which
\begin{equation}x_m^{\delta=1}=0.83815,~~~x_m^{\delta=2}=0.66939,~~~x_m^{\delta=3}=0.67755 \end{equation} to plot intensity of
the images amplification factor $|F|=\sqrt{FF^*}$ versus the
frequency $w$ in figures 5 for (\ref{FPL}) and (\ref{amdel})
respectively which are given by
\begin{equation}|F_{\delta=0}|=2|J_0(w|\vec{y}+\vec{z}|)|\end{equation}
and
\begin{equation}|F_{\delta=1,2,3,\cdots}|=\{1+2x_m\sqrt{w(1+\delta)}+2x_m^2(1+w(1+\delta))+\end{equation}$$
2x_m^3\sqrt{w(1+\delta)}[1+w(1+\delta)]+x_m^4[1+w^2(1+\delta)^2]\}^\frac{1}{2}/(1+\delta).$$
They show raising of the image amplifications factor by increasing
$w$ and $x_m$ for non point mass $\delta=1,2,3$ but not for point
 mass $\delta=0.$ In the subsequent
section we investigate how can produce the stationary images via
interference of waves and geometric optics approaches?
\subsection{Images production via interference of waves}
 At first step we substitute the lens potentials (\ref{psi0}), (\ref{psi1}), (\ref{psi2}) and (\ref{psi3}) into the lens equation
  (\ref{19}) and then we obtain images positions in the geometric optics approach. To do so we consider just highest order singular terms
  in the lens potentials. Also we assume that position vector direction of the source and the images are aligned which means the image and the object
  are on the same page  and source plane, image plane and lens plane are parallel with each others. In this sense the lens equations
  takes a simpler form
  such that
\begin{equation}\label{L0}x-y-z-\frac{1}{x}=0\end{equation}
for $\delta=0$
\begin{equation}\label{L1}x-y-z-\frac{\pi}{p}\sqrt{\frac{q(1-p)}{2p}}\frac{1}{x^2}=0
\end{equation}
for $\delta=1$
\begin{equation}\label{L2}x-y-z-\frac{q(1-p^2)}{2p^3}\frac{1}{x^3}=0
\end{equation}
for $\delta=2$ and
\begin{equation}\label{L3}x-y-z-\frac{3\pi}{4p^3}\bigg(\frac{q(1-p)}{2p}\bigg)^\frac{3}{2}\frac{1}{x^4}=0\end{equation}
for $\delta=3$ and so on. To obtain how many images are formed on
the observer plane at its center $z=0$, we plot diagrams of the
above lens equations for ansatz (\ref{ansatz}) in figures 6.
Looking at these figures one infer that for $\delta=0,2$ there are
two different images at each side of the center of the telescope
convex lens but not for $\delta=1,3.$ In the latter case just one
image is at say right side while three images at left  side. This
shows that distribution of images on the observer plane are
dependent to shape of  the  mass function of the lens. In other
word if numeric value in $\delta$ parameter become even (odd)
number then number of images will be (unequal) equal in both side
of center of the convex lens of the telescope. Magnification of
the images in the geometric optics limit is defined by
$\Upsilon=\big(\frac{y}{x}\frac{dy}{dx}\big)^{-1}$ which by
substituting (\ref{L0}), (\ref{L1}), (\ref{L2}) and (\ref{L3}) we
plotted its absolute value in figures 7. They show inequality
between magnifications of left side images and right side ones
when $\delta=1,3$ and more precisely that for left side images
  the magnifications are greater
than that of ones for right side images.
 To produce stationary images from interference of fringes pattern given by (\ref{36}) we should calculate images position
 $x_i$ and corresponding magnifications  (\ref{Ups}) for a fixed source position $y$ by applying the lens equations (\ref{L0}),
  (\ref{L1}), (\ref{L2}) and (\ref{L3}).  Then we substitute them into the (\ref{36}). We do these
   for  sample source positions $y=0,3,-3$ and collect them in the table 1.
\begin{center}
Table 1. Stationary images position and corresponding
magnifications for $\delta=0,1,2,3$ and
$p=\frac{1}{2},q=\frac{1}{10}$ at $z=0$
 \end{center}
\begin{center}
\begin{tabular}{|c|c|c|c|c|}
  \hline
 $\delta$& $y$ & $(x_1,|\Upsilon_1|)$ & $(x_2,|\Upsilon_2|)$ & $(x_3,|\Upsilon_3|)$ \\
  \hline
$0$ & -3 & (+0.30,~0.008) & (-3.30,~1.008) & --- \\
$0$ &0 & (+1,~$\infty$) & (-1,~$\infty$) & --- \\
 0&  +3 & (-0.30,~0.008) & (+3.30,~1.008) & --- \\
\hline
$1$ & -3 & (-2.92,~1.03) & (-0.53,~0.02) & (+0.45,~0.009) \\
$1$ &0 & (+0.89,~$\infty$) & --- & --- \\
1&  +3 & (+3.07,0.98) & --- & --- \\
\hline
$2$ & -3 & (-3.00,~1.00) & (+0.12,~0.001) & --- \\
$2$ &0 & (-0.26,~3.75) & (+0.26,~3.75) & --- \\
 2&  +3 & (-0.12,0.001) & (+3.00,1.00) & --- \\
\hline
$3$ & -3 & (-2.99,~1.00) & (-0.54,~0.01) & (+0.50,~0.01) \\
$3$ &0 & (+0.73,~8429.4) & --- & --- \\
 3&  +3 & (3.00,~1.00) & --- & --- \\
  \hline
\end{tabular}
\end{center}
 For $p=\frac{1}{2}$ and $q=\frac{1}{10}$ the equation (\ref{36}) for each stationary image position $x_i$ become
  \begin{equation}\label{phiQ}\phi_y^{\delta}(z)=\sum_{i=1}^n|\Upsilon_i|\exp\{(iw/2)[y^2+Qz^2+5+|\vec{x}_i-\vec{y}-\vec{z}|^2\end{equation}
  $$-2\Psi_{\delta}(x_i)]\}\frac{J_1(wR|\vec{x}_i+Q\vec{z}|)}{wR|\vec{x}_i+Q\vec{z}|
  }$$
for which \begin{equation}Q=\frac{D_{LO}}{pf}>>1\end{equation} and
\begin{equation}\Psi_{\delta=0}=\ln
x,~~~\Psi_{\delta=1}=\frac{-0.7}{x},~~~\Psi_{\delta=2}=\frac{-0.15}{x^2},~~~\Psi_{\delta=3}=\frac{-0.04}{x^3}.\end{equation}
Because image plane and source plane and lens plane are parallel
with each other then we can assume $\vec{x}_i,\vec{y},\vec{z}$ are
parallel and so the equations (\ref{phiQ}) approaches to the
following form.
\begin{equation}\phi_{-3}^0(z)=0.008\exp\{(iw/2)[27.3+(Q+1)z^2-6.6z]\}\frac{J_1(wR(Qz+0.3))}{wR(Qz+0.3)}\end{equation}
$$+1.008\exp\{(iw/2)[11.7+(Q+1)z^2+0.6z]\}\frac{J_1(wR(Qz-3.3))}{wR(Qz-3.3)}$$
for $\delta=0,y=-3,$
\begin{equation}\phi_{0}^0(z)=\infty\exp\{(iw/2)[7.4+(Q+1)z^2-2z]\}\frac{J_1(wR(Qz+1))}{wR(Qz+1)}\end{equation}
$$+\infty\exp\{(iw/2)[4.6+(Q+1)z^2+2z]\}\frac{J_1(wR(Qz-1))}{wR(Qz-1)}$$
for $\delta=0,y=0,$
\begin{equation}\phi_{3}^0(z)=0.008\exp\{(iw/2)[27.3+(Q+1)z^2+6.6z]\}\frac{J_1(wR(Qz-0.3))}{wR(Qz-0.3)}\end{equation}
$$+1.008\exp\{(iw/2)[11.7+(Q+1)z^2-0.6z]\}\frac{J_1(wR(Qz+3.3))}{wR(Qz+3.3)}$$
for $\delta=0,y=3,$
\begin{equation}\phi_{-3}^1(z)=1.03\exp\{(iw/2)[13.53+(Q+1)z^2-0.16z]\}\frac{J_1(wR(Qz-2.92))}{wR(Qz-2.92)}\end{equation}
$$+0.02\exp\{(iw/2)[17.46+(Q+1)z^2-4.94z]\}\frac{J_1(wR(Qz-0.53))}{wR(Qz-0.53)}$$$$+0.009\exp\{(iw/2)[29.01+(Q+1)z^2-6.9
z]\}\frac{J_1(wR(Qz+0.45))}{wR(Qz+0.45)}$$ for $\delta=1,y=-3,$
\begin{equation}\phi_{0}^1(z)=\infty\exp\{(iw/2)[7.36+(Q+1)z^2-1.78z]\}\frac{J_1(wR(Qz+0.89))}{wR(Qz+0.89)}\end{equation}
for $\delta=1,y=0$
\begin{equation}\phi_{3}^1(z)=0.98\exp\{(iw/2)[14.46+(Q+1)z^2-0.14z]\}\frac{J_1(wR(Qz+3.07))}{wR(Qz+3.07)}\end{equation}
for $\delta=1,y=3$
\begin{equation}\phi_{-3}^2(z)=1.00\exp\{(iw/2)[50.03+(Q+1)z^2+12z]\}\frac{J_1(wR(Qz-3))}{wR(Qz-3)}\end{equation}
$$+0.001\exp\{(iw/2)[43.13+(Q+1)z^2+5.76z]\}\frac{J_1(wR(Qz+0.12))}{wR(Qz+0.12)}$$
for $\delta=2,y=-3,$
\begin{equation}\phi_{0}^2(z)=3.75\exp\{(iw/2)[9.51+(Q+1)z^2+0.52z]\}\frac{J_1(wR(Qz-0.26))}{wR(Qz-0.26)}\end{equation}
$$+3.75\exp\{(iw/2)[9.51+(Q+1)z^2-0.52z]\}\frac{J_1(wR(Qz+0.26))}{wR(Qz+0.26)}$$
for $\delta=2,y=0,$
\begin{equation}\phi_{3}^2(z)=0.001\exp\{(iw/2)[44.57+(Q+1)z^2+6.24z]\}\frac{J_1(wR(Qz-0.12))}{wR(Qz-0.12)}\end{equation}
$$+1.00\exp\{(iw/2)[14.03+(Q+1)z^2]\}\frac{J_1(wR(Qz+3))}{wR(Qz+3)}$$
for $\delta=2,y=3,$
\begin{equation}\phi_{-3}^3(z)=1.00\exp\{(iw/2)[14+(Q+1)z^2-0.02z]\}\frac{J_1(wR(Qz-2.99))}{wR(Qz-2.99)}\end{equation}
$$+0.01\exp\{(iw/2)[19.54+(Q+1)z^2-4.92z]\}\frac{J_1(wR(Qz-0.54))}{wR(Qz-0.54)}$$$$+0.01\exp\{(iw/2)[26.89+(Q+1)z^2-7.0
z]\}\frac{J_1(wR(Qz+0.5))}{wR(Qz+0.5)}$$ for $\delta=3,y=-3,$
\begin{equation}\phi_{0}^3(z)=8429.4\exp\{(iw/2)[5.74+(Q+1)z^2-1.46z]\}\frac{J_1(wR(Qz+0.73))}{wR(Qz+0.73)}\end{equation}
for $\delta=3,y=0$ and
\begin{equation}\phi_{3}^3(z)=1.00\exp\{(iw/2)[14.00+(Q+1)z^2]\}\frac{J_1(wR(Qz+3))}{wR(Qz+3)}\end{equation}
for $\delta=3,y=3.$ We call intensity of the above waves by
\begin{equation}\label{intensityy}I_{y}^{\delta}(z)=\sqrt{\phi_y^\delta(z)\phi_y^{\delta *}(z)}\end{equation}
in which $\phi_y^{\delta *}(z)$ is complex conjugate of the wave
$\phi_y^{\delta}(z)$ and their diagrams are plotted versus $z$ in
the observer plane in figure 8. in fact peaks in the interference
pattern of the waves in this figures show number and location of
images. All diagrams in the figure 8 are plotted for particular
initial values
\begin{equation}Q=1000,~~~w=100,~~~R=0.5,~~~\infty\to1.\end{equation}
Peaks of waves in figures 8 are comparable with number of
geometric images in figure 6. Also height of diagrams describe
value of brightness of the images. The analysis and interpretation
of the results of this work are postponed to the next section.
  \section{Concluding remark and outlook}
In this work we studied gravitational lensing  in both geometrical
and waves optics approaches for  a spherically symmetric static
black hole which asymptotically behaves as modified Schwarzschild
de Sitter black hole  in weak field limits. Modifications are
logarithmic metric potential and mass distribution function
instead of the point mass. They are generated from time like
vector field which is coupled non minimally with the well known
scalar tensor Brans Dicke gravity. We obtained that this black
hole has two event horizons for positive numeric cosmological
parameter while it has just one event horizon for negative
cosmological parameter. By fixing numeric value of the positive
cosmological parameter there is obtained a particular value for
the vector field parameter for which two horizons of the black
hole get closer to each other. By considering this black hole
metric to be lens we studied gravitational lensing of light of a
point star far from the lens and the observer. To do this we
obtained position of the stationary images and corresponding
magnifications in the geometric optics limit by solving the lens
equation. Instead of light rays in geometric optics limit we used
Fresnel-Kirchhoff diffraction theory of massless scalar waves to
study the gravitational lensing via wave optics limit and we
obtained location of stationary images by interference of fringes.
Lens potential of our model has singular term at $x\to0$ as
$1/x^\delta$ versus the image position $x$ in which the parameter
$\delta$ is related to the radial component of the time like
vector field. Mathematical derivations show that for  numeric even
values of the $\delta$ parameter ($\delta=0,2,\cdots$) number of
images are equal in left side and right side of the center of
local cartesian coordinates on the observer plane (eyepiece of
telescope) but not for odd numeric values of the $\delta$
parameter ($\delta=1,3,\cdots$). If we want to say in precise,
there is just one image in left side and one image in right side
of center of the coordinates for $\delta=0,2,\cdots.$ While for
numeric odd value of $\delta,$ images numbers is depended to
source position namely if the source to be at right side of the
optical axis $(y>0)$ then there is just one magnified image in
right side of the coordinate center on the observer plane $(z>0)$
but for sources which is located at left side of the optical axis
then we will have three images which one of them is located in
right side of the coordinate center and two of them are located in
left side of the coordinate center on the observer plane (eyepiece
of telescope). Magnification of nearest image to the optical axis
has highest intensity of brightness and the furthest image to the
optical axis has minimum intensity of brightness. In the geometric
optics limits there is a single Einstein`s ring if the source and
the lens and the image to be at straightforward line (the cases
$y=0$ in the diagrams of figures 6 and 8) but in the wave optics
limit there are more Einstein`s ring with same center because of
the interference of fringes. In the geometric optics limit we see
that center of the Einstein`s ring is dark but not in the wave
optics limit. In the latter case center of the rings have some
small brightness. As an application of the gravitational lensing
by wave optics limits we like to investigate black hole shadow and
gravitational lensing of the gravitational waves instead of the
electromagnetic waves as our future works. One of important
applications of the wave optics gravitational lensing in the
astrophysics is determination of size of the black holes by study
on interference fringes of diffracted electromagnetic waves. Hence
study of polarization effects of the electromagnetic waves on the
gravitational lensing is also useful work which one can consider
to do as extension of the present work.
\\
\\
\textbf{Acknowledgments}\\
 This work was supported in part by the
Semnan University Grant No.1398-07-061108 for Scientific
Research.\\

\begin{figure}[ht]
\centering  \subfigure[{}]{\label{1}
\includegraphics[width=0.45\textwidth]{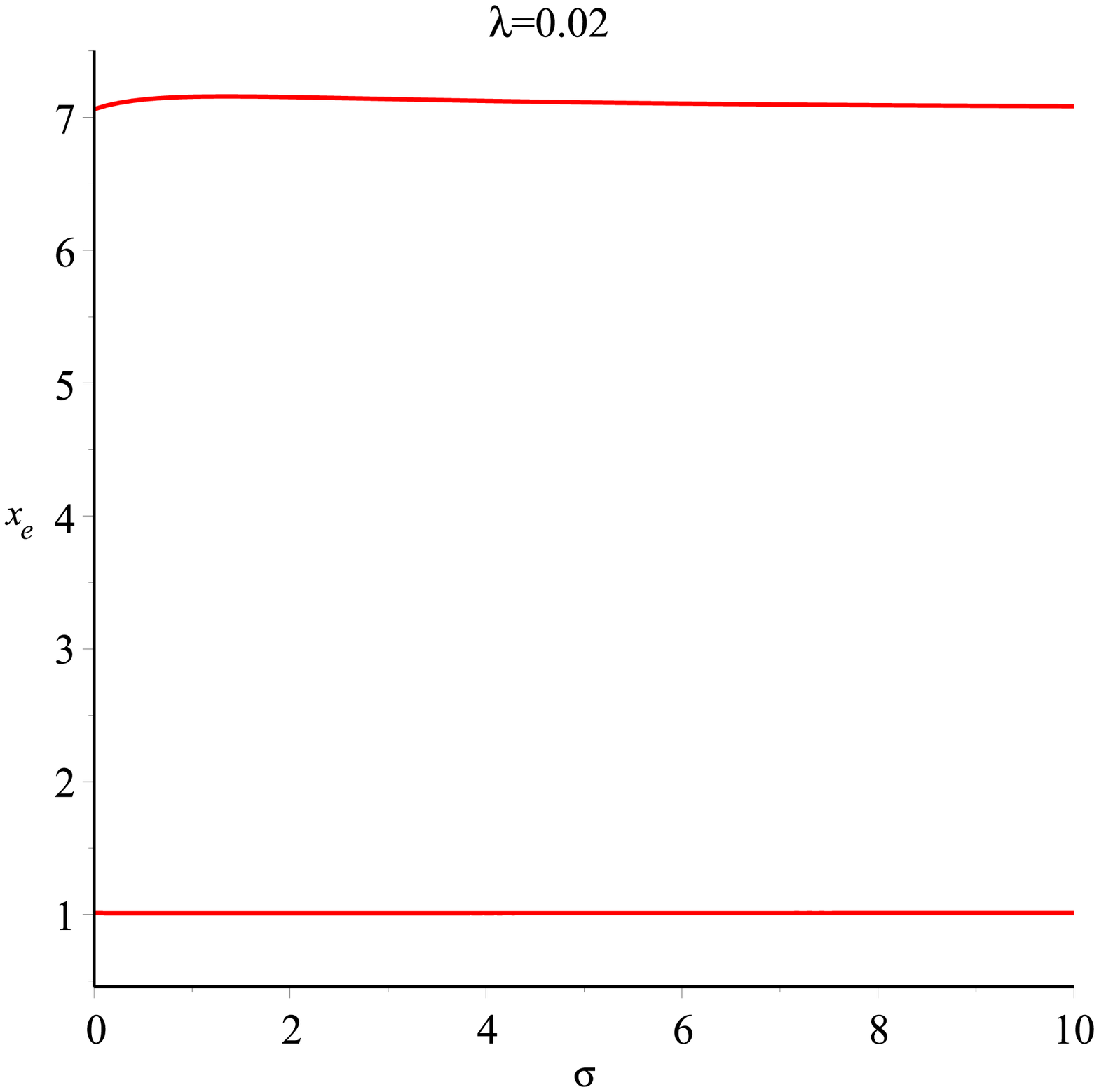}}
\hspace{3mm} \subfigure[{}]{\label{1}
\includegraphics[width=0.45\textwidth]{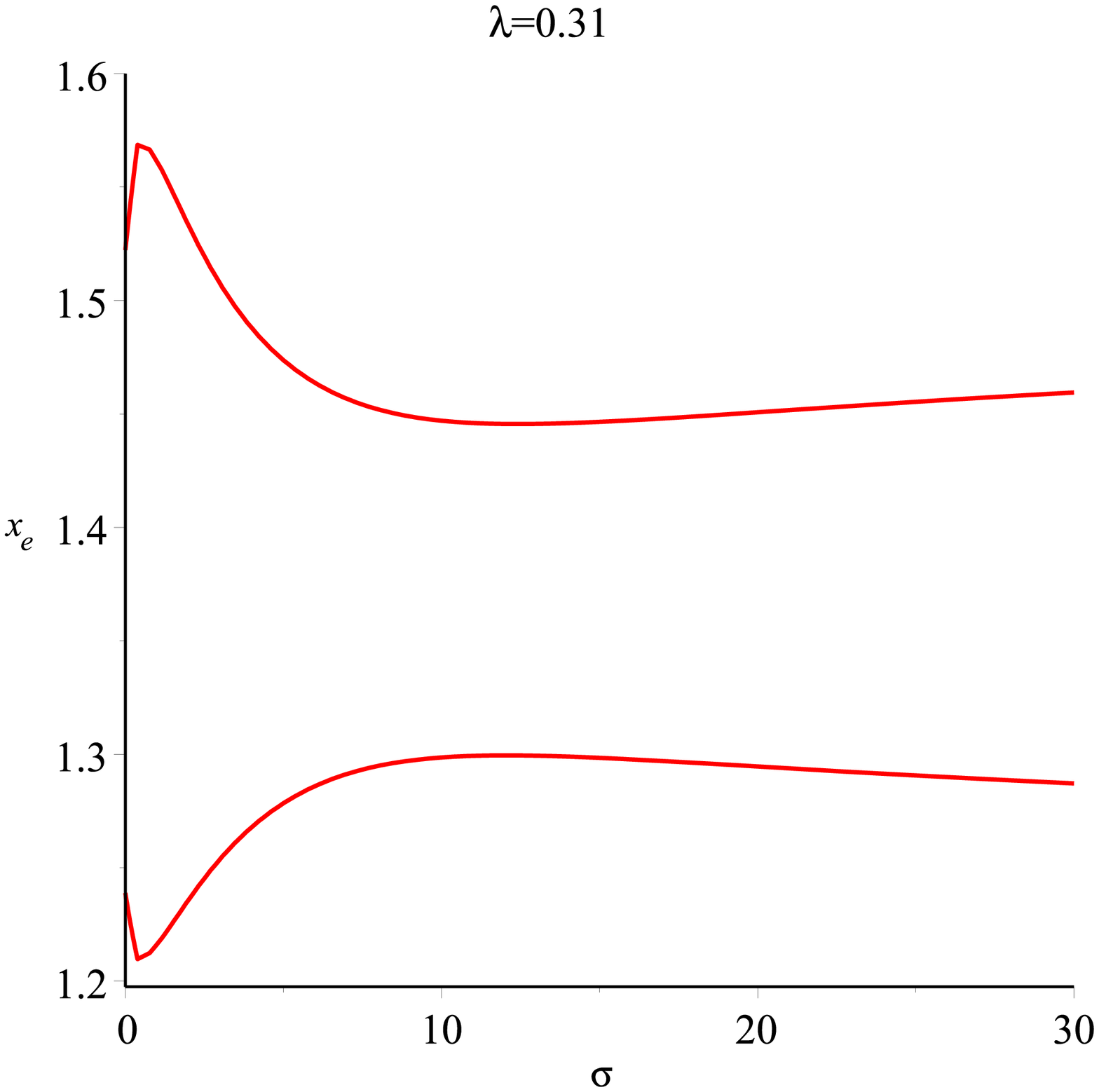}}
\hspace{3mm} \subfigure[{}]{\label{1}
\includegraphics[width=0.45\textwidth]{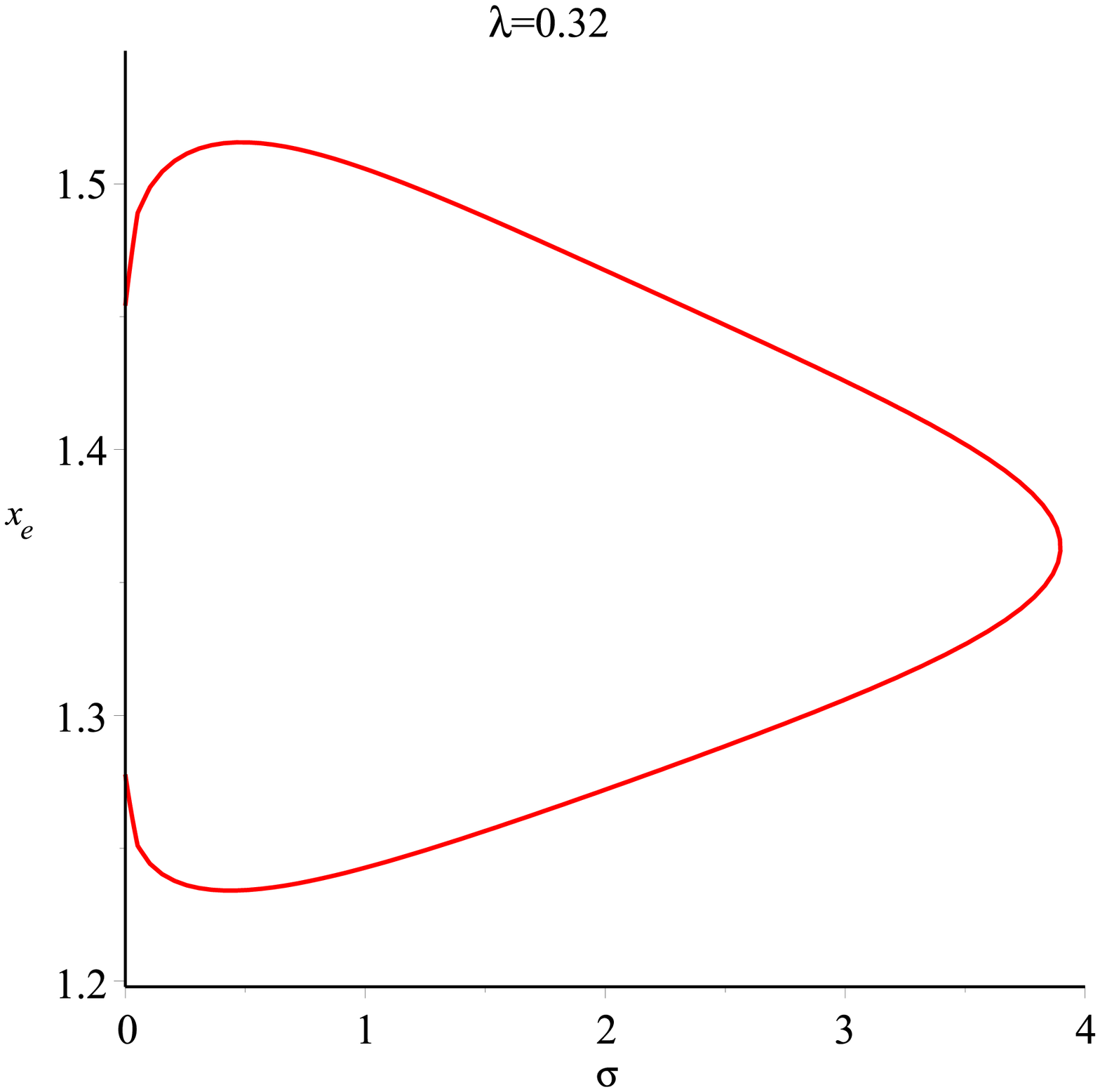}}
\hspace{3mm} \subfigure[{}]{\label{1}
\includegraphics[width=0.45\textwidth]{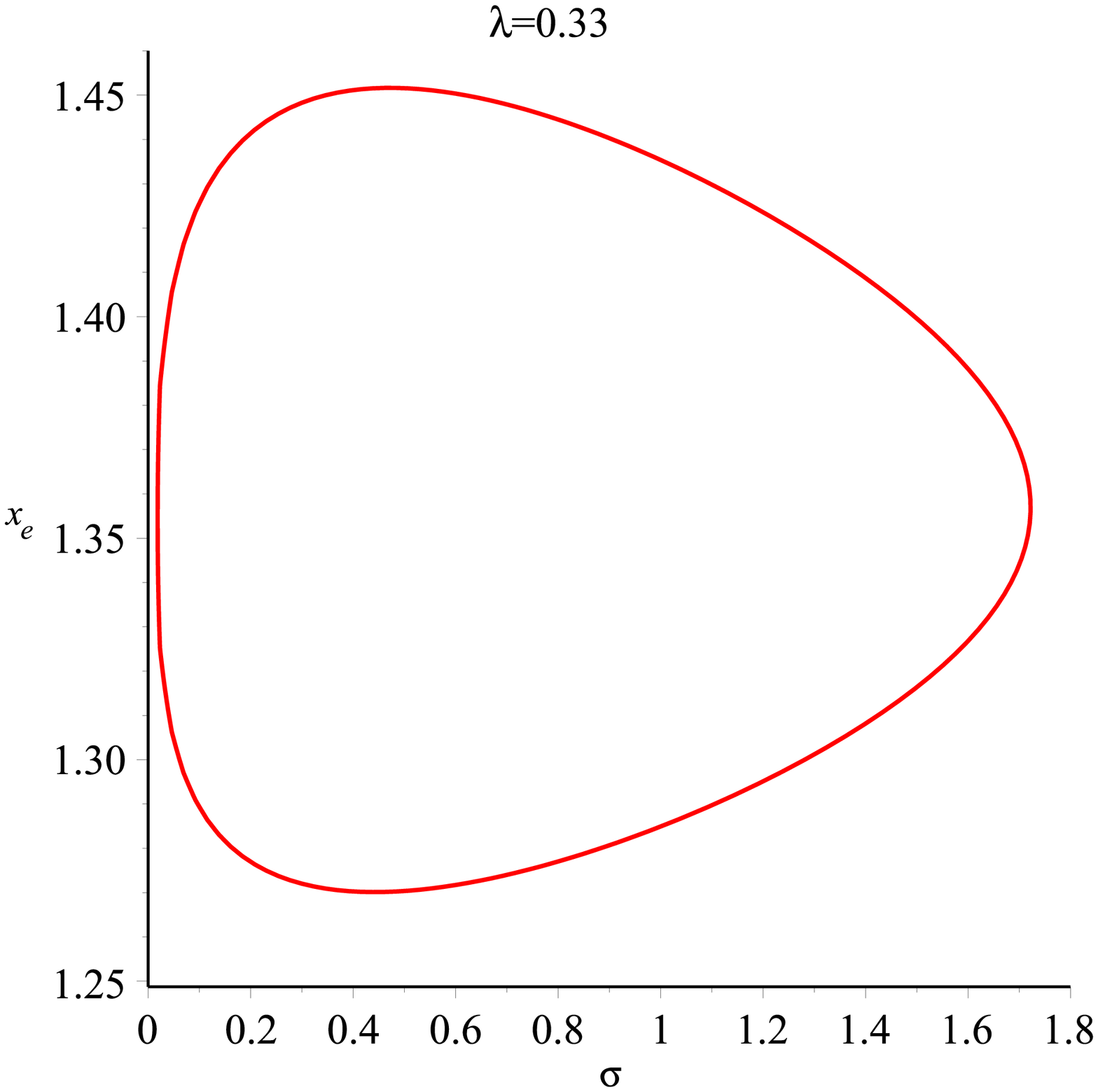}}
\hspace{3mm}
 \caption{Diagrams of the black hole event horizons $x_e$ versus observer acceleration parameter $\sigma$ for  dimensionless cosmological parameter
 $0.2<\lambda<0.33.$ There are not obtained other diagrams for
 $\lambda<0.2$ and $\lambda>0.33$ }
\end{figure}
\begin{figure}[ht]
\centering  \subfigure[{}]{\label{1}
\includegraphics[width=0.45\textwidth]{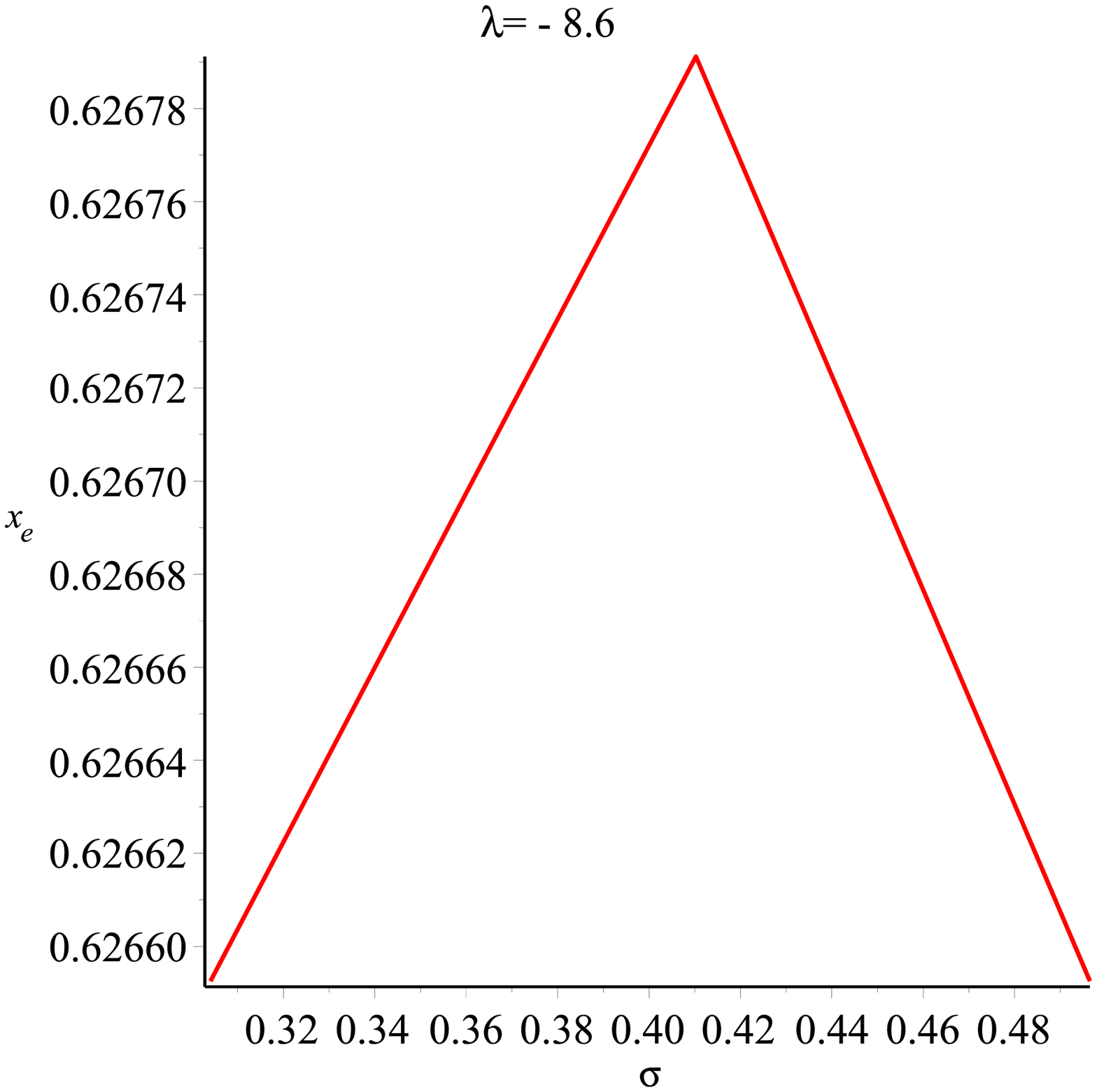}}
\hspace{3mm} \subfigure[{}]{\label{1}
\includegraphics[width=0.45\textwidth]{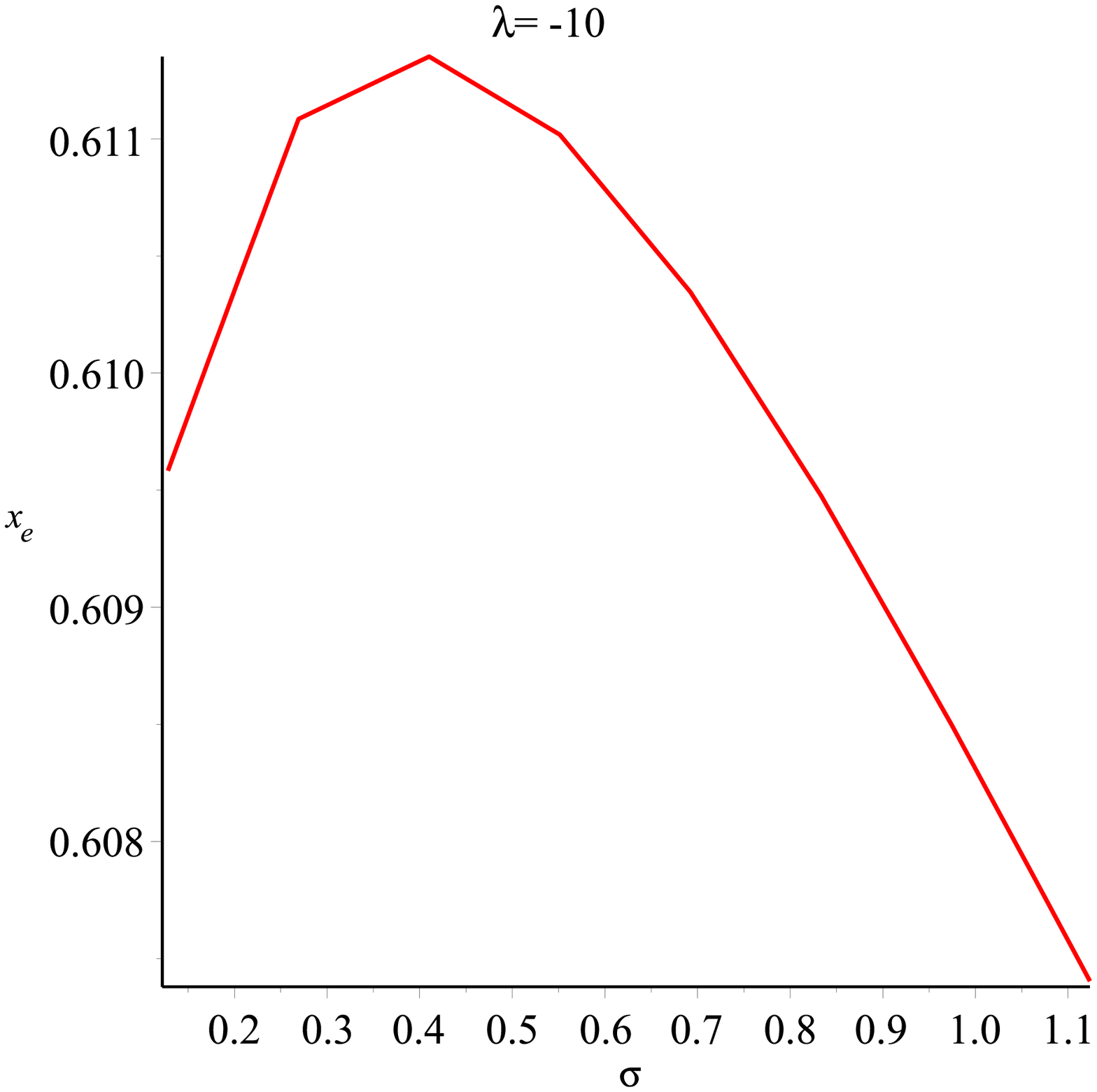}}
\hspace{3mm} \subfigure[{}]{\label{1}
\includegraphics[width=0.45\textwidth]{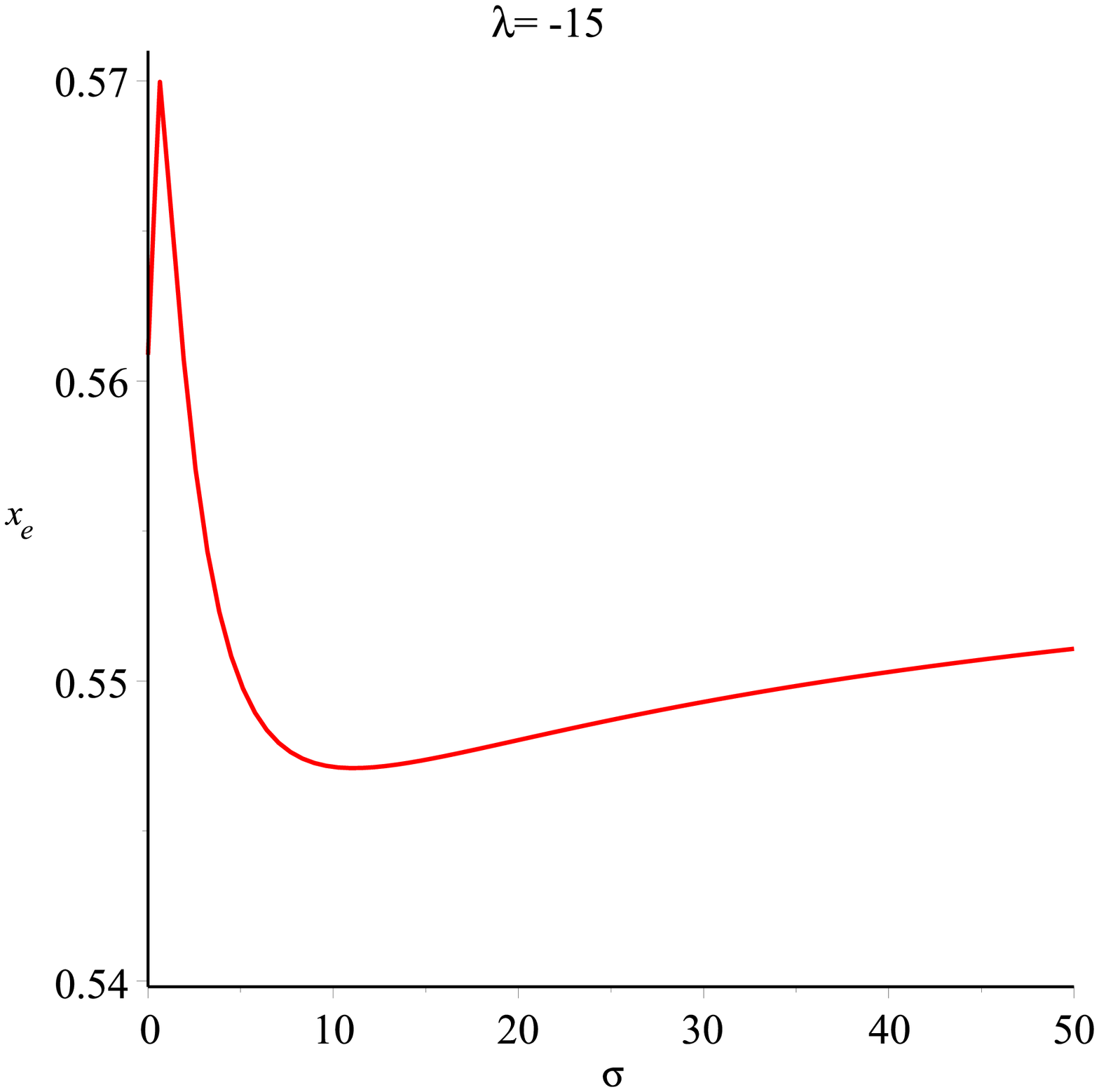}}
\hspace{3mm} \subfigure[{}]{\label{1}
\includegraphics[width=0.45\textwidth]{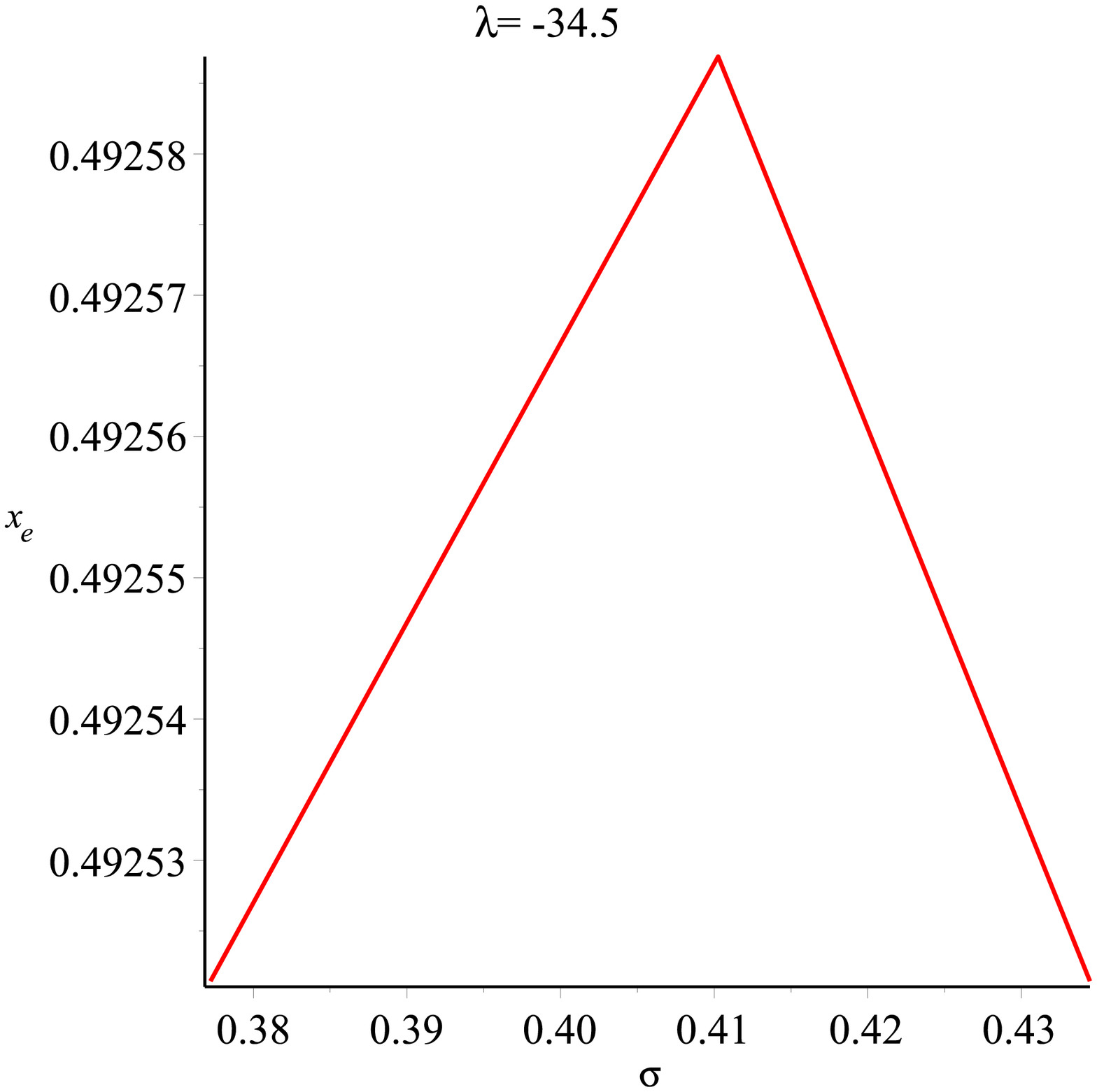}}
\hspace{3mm}
 \caption{Diagrams of the black hole event horizons $x_e$ versus observer acceleration parameter $\sigma$ for  dimensionless cosmological parameter
 $-34.5<\lambda<-8.6.$ There are not obtained other diagrams for
 $\lambda<-34.5$ and $\lambda>-0.86$}
\end{figure}
\begin{figure}[ht]
\centering  \subfigure[{}]{\label{1}
\includegraphics[width=0.45\textwidth]{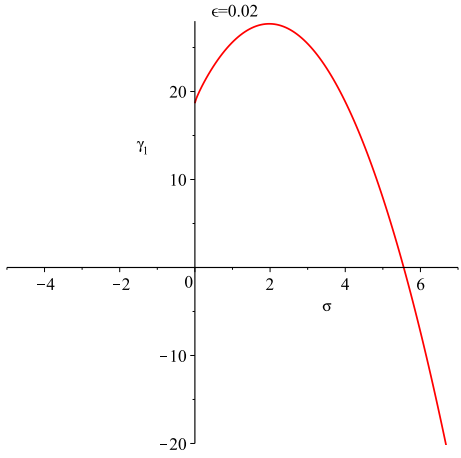}}
\hspace{3mm} \subfigure[{}]{\label{1}
\includegraphics[width=0.45\textwidth]{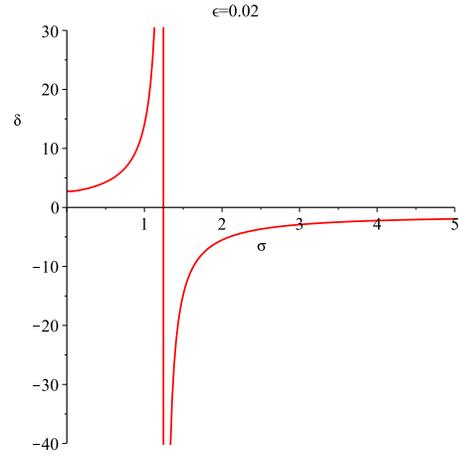}}
\hspace{3mm} \subfigure[{}]{\label{1}
\includegraphics[width=0.45\textwidth]{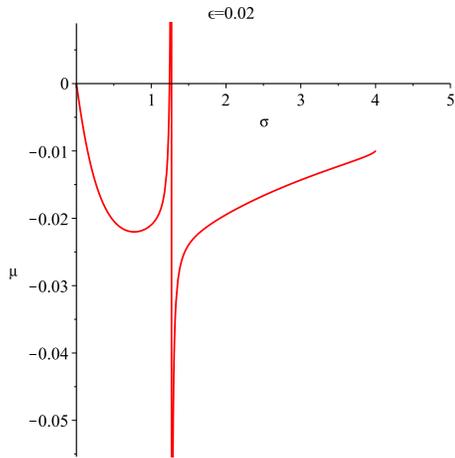}}
\hspace{3mm} \subfigure[{}]{\label{1}
\includegraphics[width=0.45\textwidth]{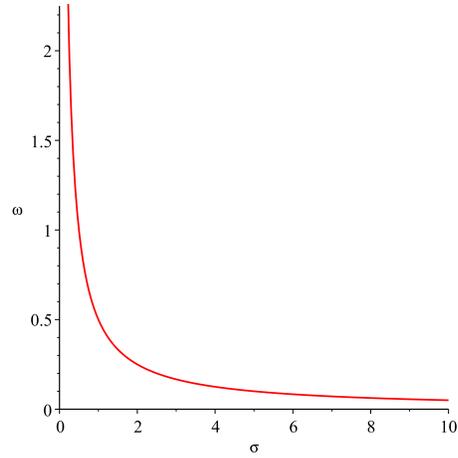}}
\hspace{3mm}
 \caption{Numeric values for parameters of the  scalar vector Brans Dicke cosmological black holes }
\end{figure}
\begin{figure}[ht]
\centering  \subfigure[{}]{\label{1}
\includegraphics[width=0.9\textwidth]{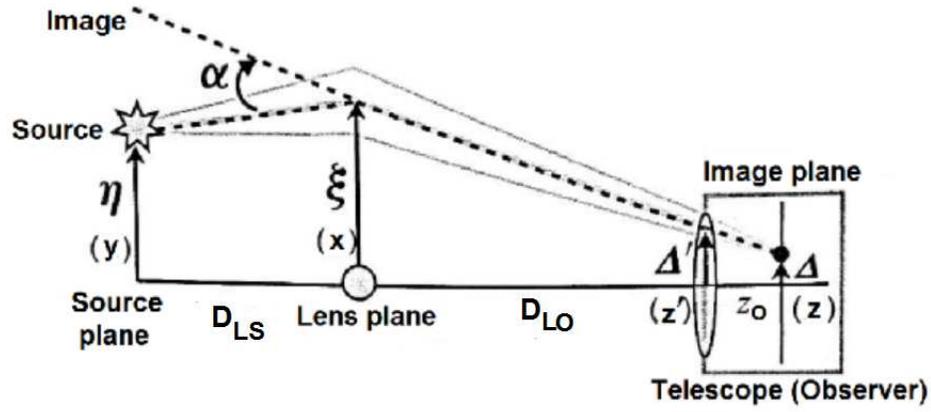}}
\hspace{3mm} \subfigure[{}]{\label{1}
\includegraphics[width=0.9\textwidth]{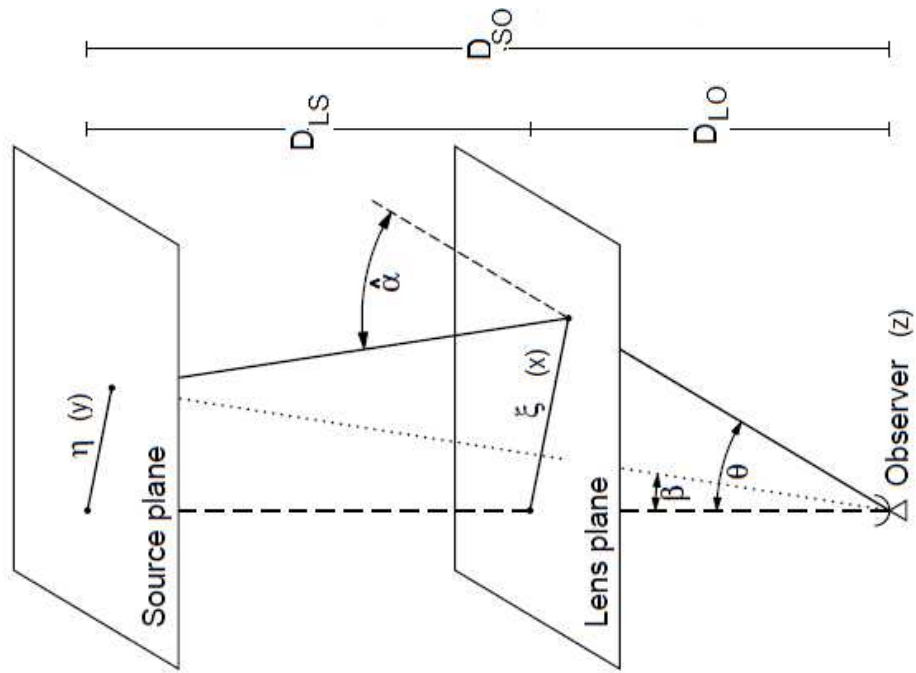}}
\hspace{3mm}
 \caption{Geometry of the gravitational lens (a) from \cite{YN} and (b) from \cite{Bar} respectively with some minor revisions }
\end{figure}
\begin{figure}[ht]
\centering  \subfigure[{}]{\label{1}
\includegraphics[width=0.45\textwidth]{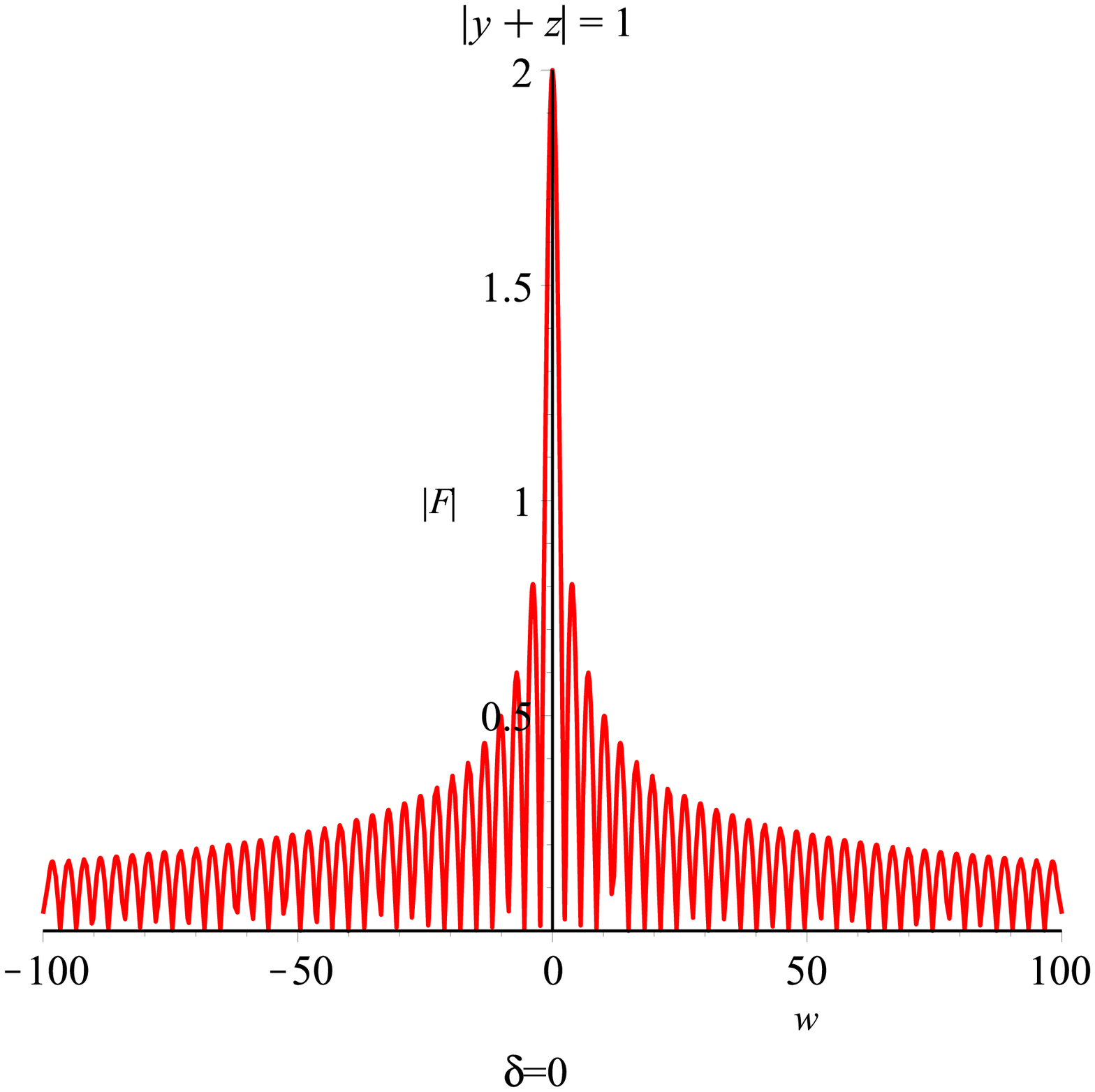}}
\hspace{3mm} \subfigure[{}]{\label{1}
\includegraphics[width=0.45\textwidth]{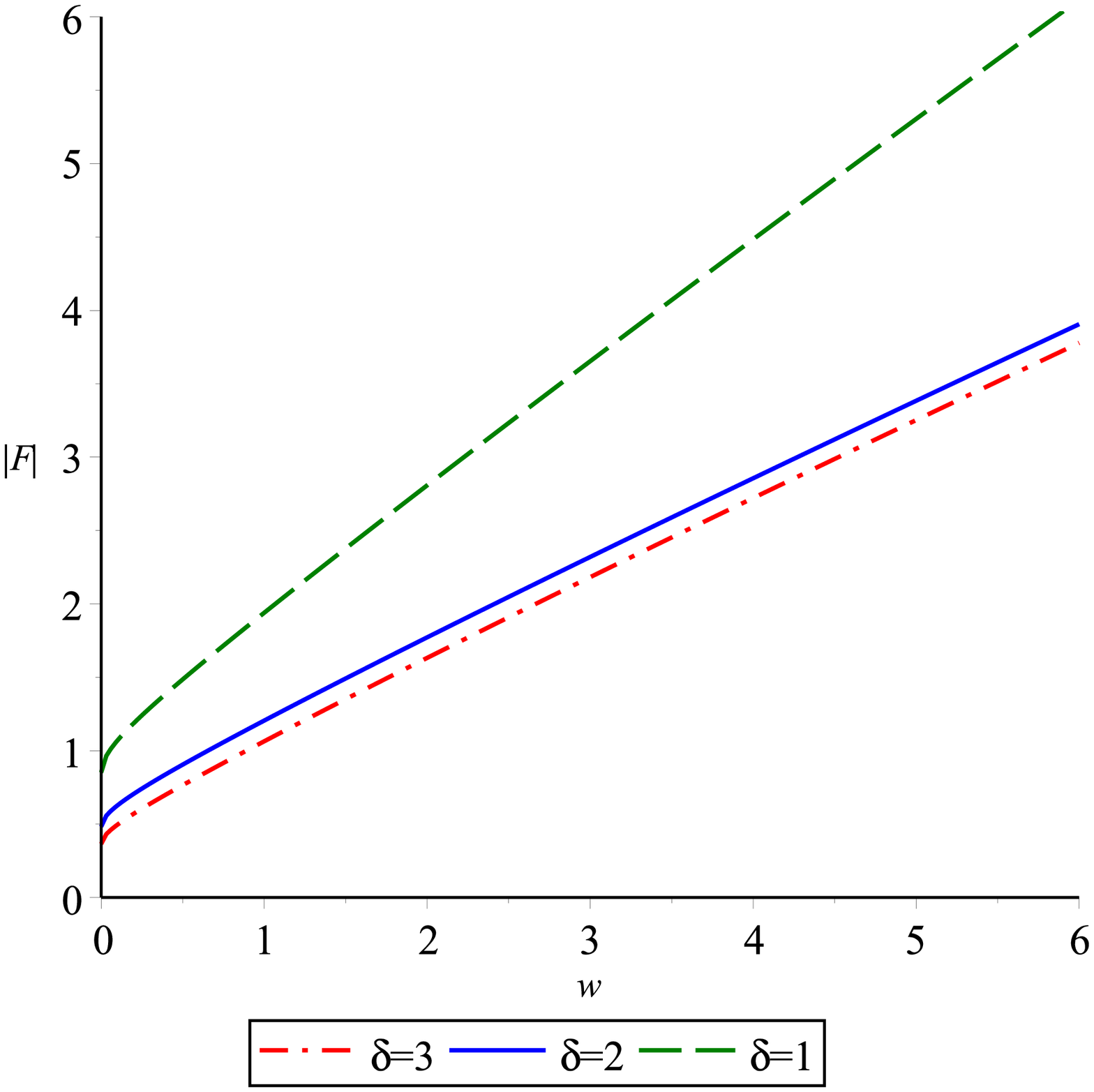}}
\hspace{3mm} \subfigure[{}]{\label{1}
\includegraphics[width=0.45\textwidth]{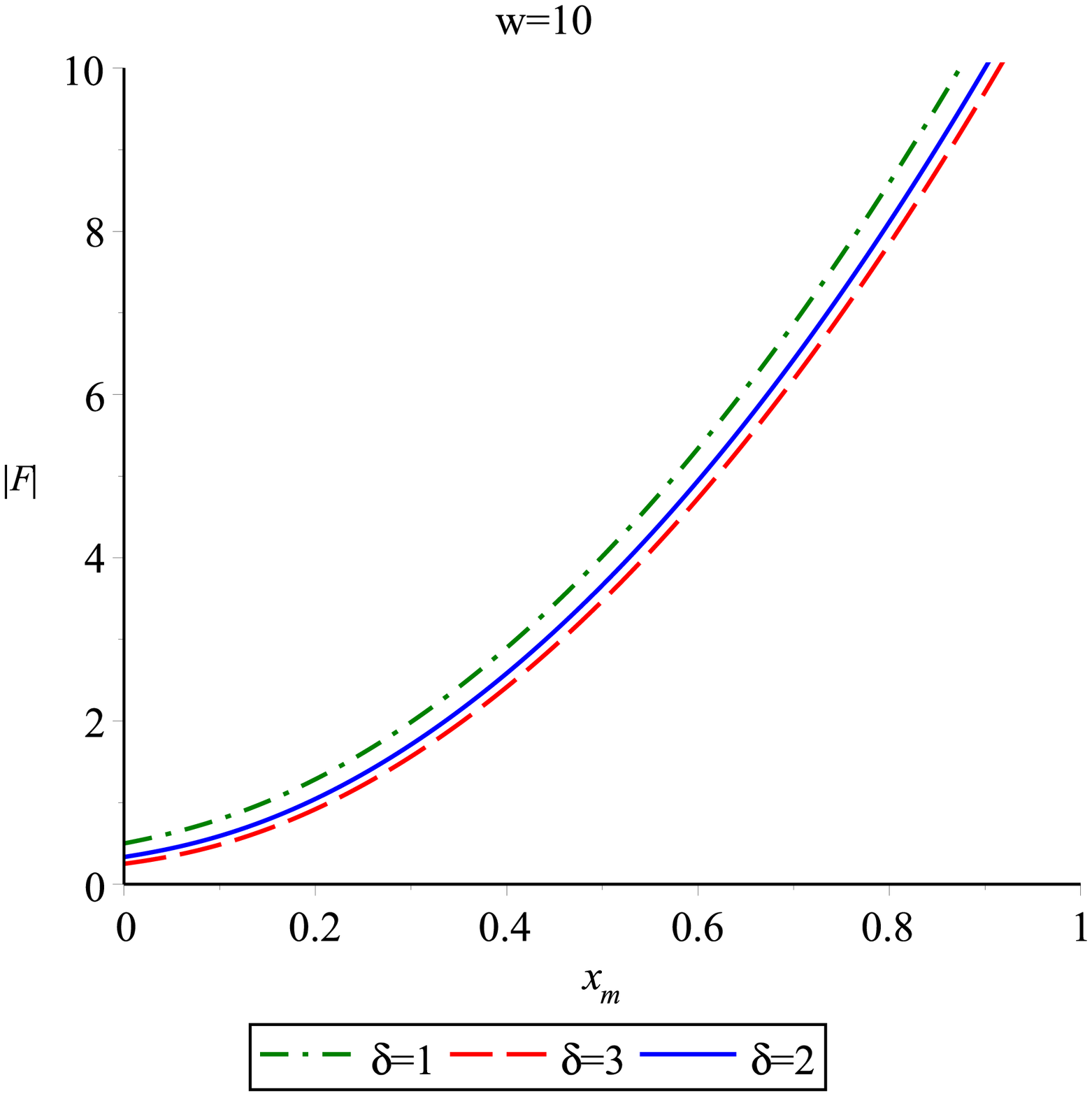}}
\hspace{3mm}
%\subfigure[{}]{\label{1}
%\includegraphics[width=0.45\textwidth]{omega}}
%\hspace{3mm}
 \caption{Amplification factor absolute value is plotted for point lens versus the dimensionless frequency
 $w$ in (a) and is plotted versus $w$ and the minimum point of non point black hole lens potentials respectively in (b) and (c).}
\end{figure}
\begin{figure}[ht]
\centering  \subfigure[{}]{\label{1}
\includegraphics[width=0.45\textwidth]{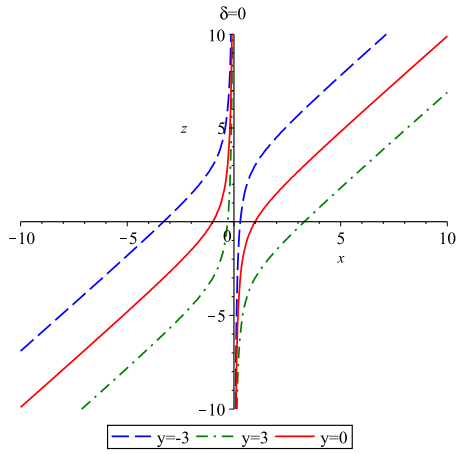}}
\hspace{3mm} \subfigure[{}]{\label{1}
\includegraphics[width=0.45\textwidth]{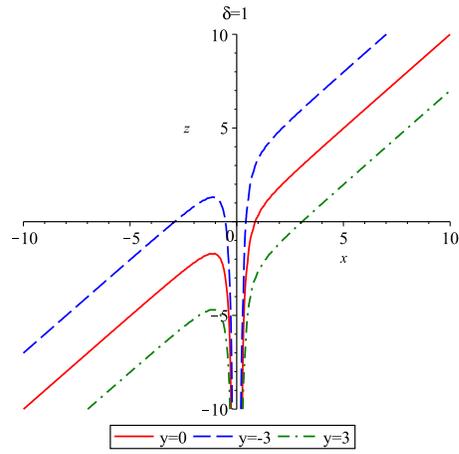}}
\hspace{3mm} \subfigure[{}]{\label{1}
\includegraphics[width=0.45\textwidth]{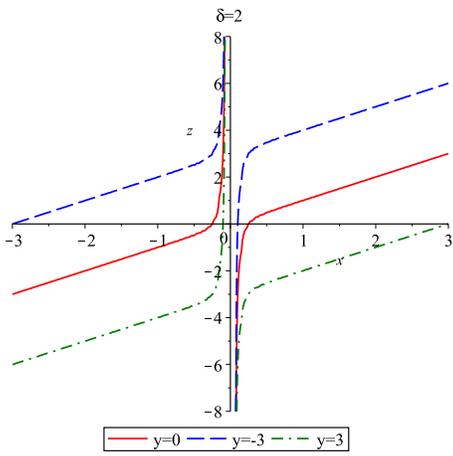}}
\hspace{3mm} \subfigure[{}]{\label{1}
\includegraphics[width=0.45\textwidth]{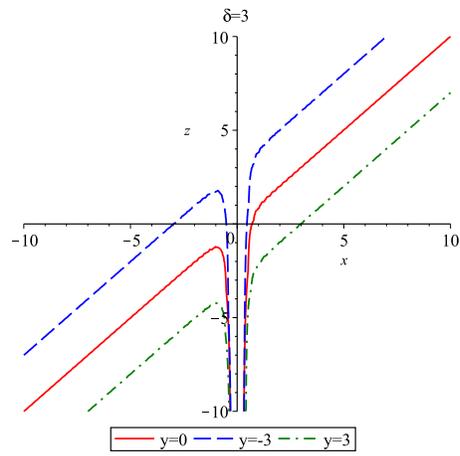}}
\hspace{3mm}
 \caption{ Images position on the observer plane in geometric optics limit }
\end{figure}
\begin{figure}[ht]
\centering  \subfigure[{}]{\label{1}
\includegraphics[width=0.45\textwidth]{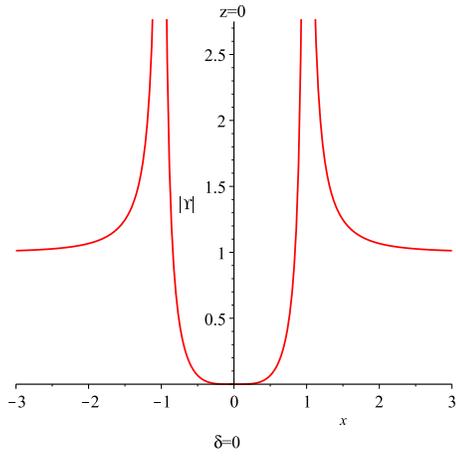}}
\hspace{3mm} \subfigure[{}]{\label{1}
\includegraphics[width=0.45\textwidth]{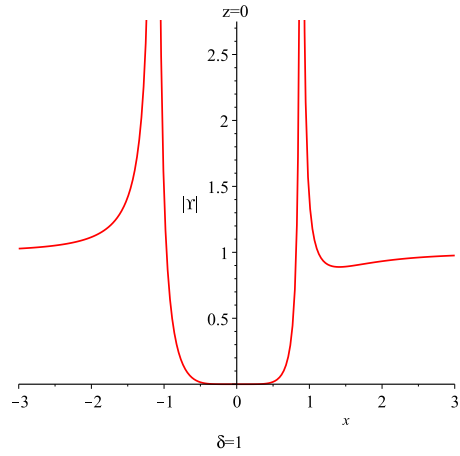}}
\hspace{3mm} \subfigure[{}]{\label{1}
\includegraphics[width=0.45\textwidth]{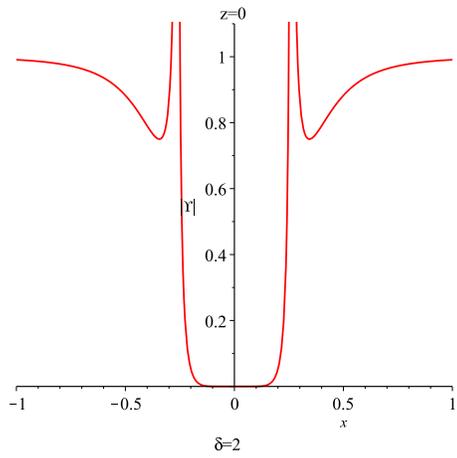}}
\hspace{3mm} \subfigure[{}]{\label{1}
\includegraphics[width=0.45\textwidth]{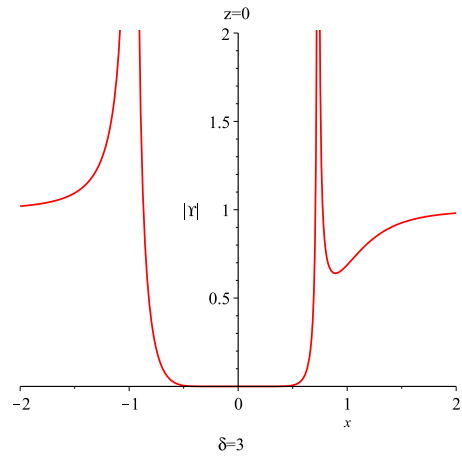}}
\hspace{3mm} \caption{ Magnification of images in geometric optics
limit }
\end{figure}
\begin{figure}[ht] \centering
 {\label{1}
\includegraphics[width=0.32\textwidth]{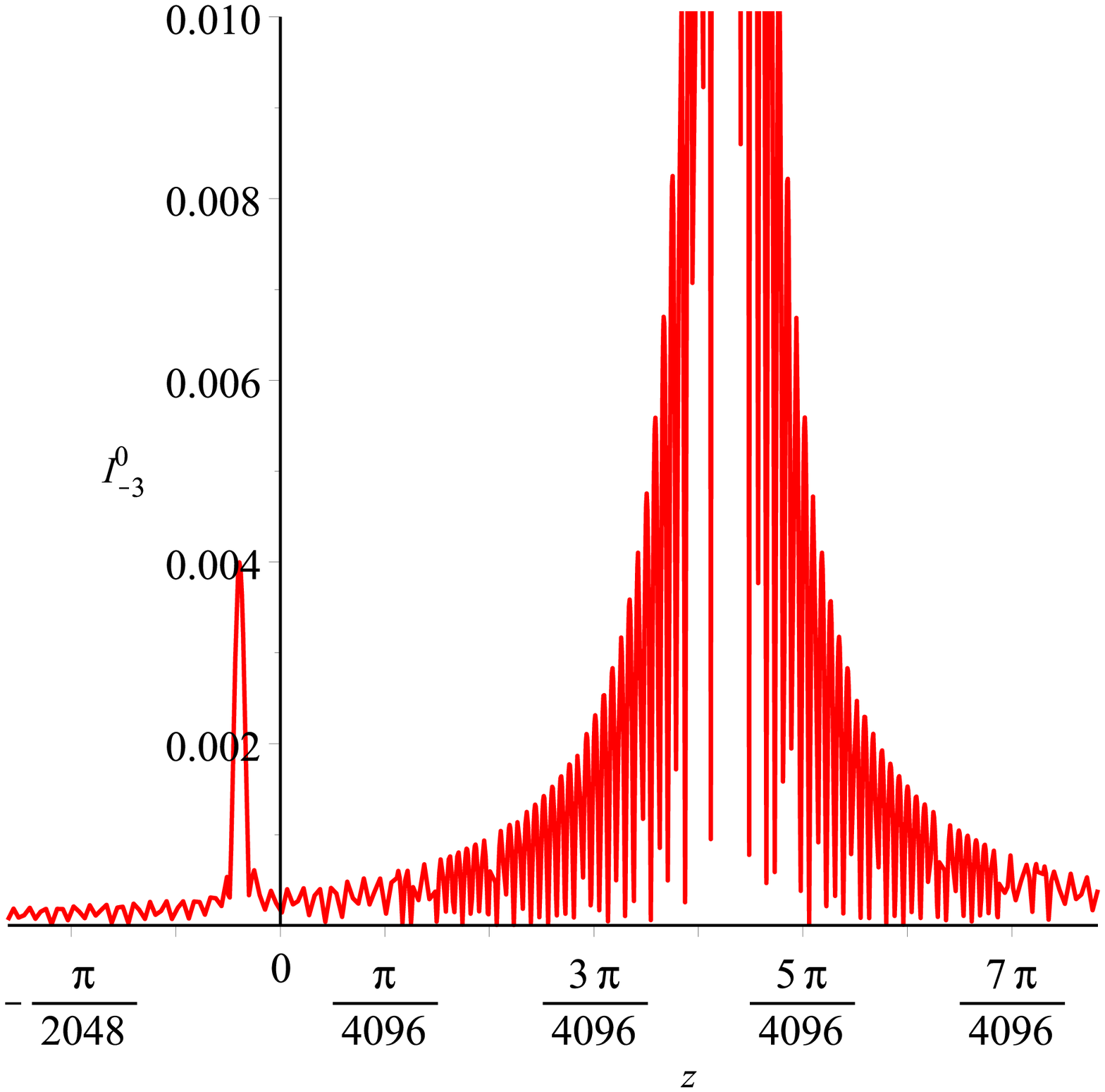}
\includegraphics[width=0.32\textwidth]{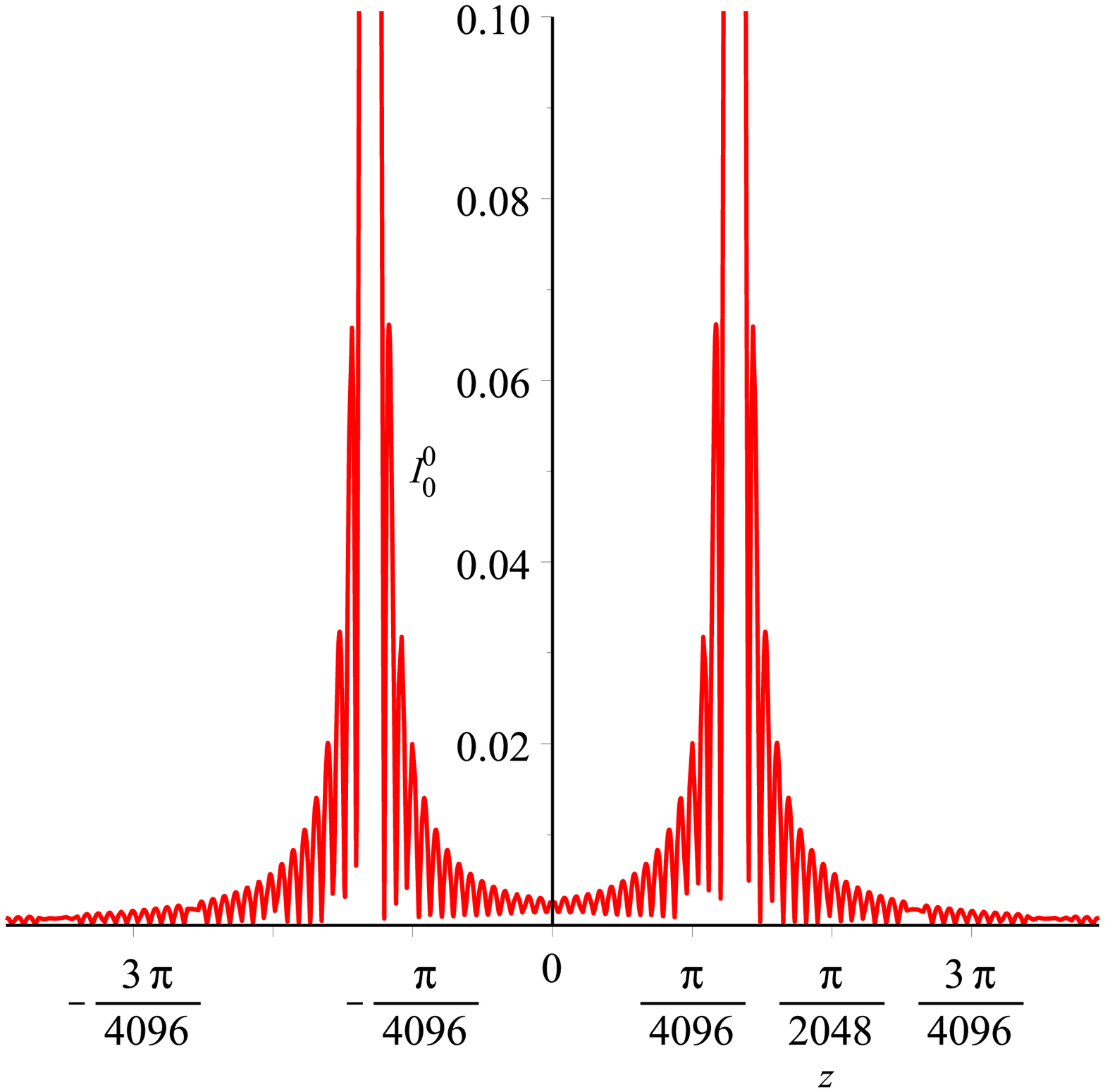}
\includegraphics[width=0.32\textwidth]{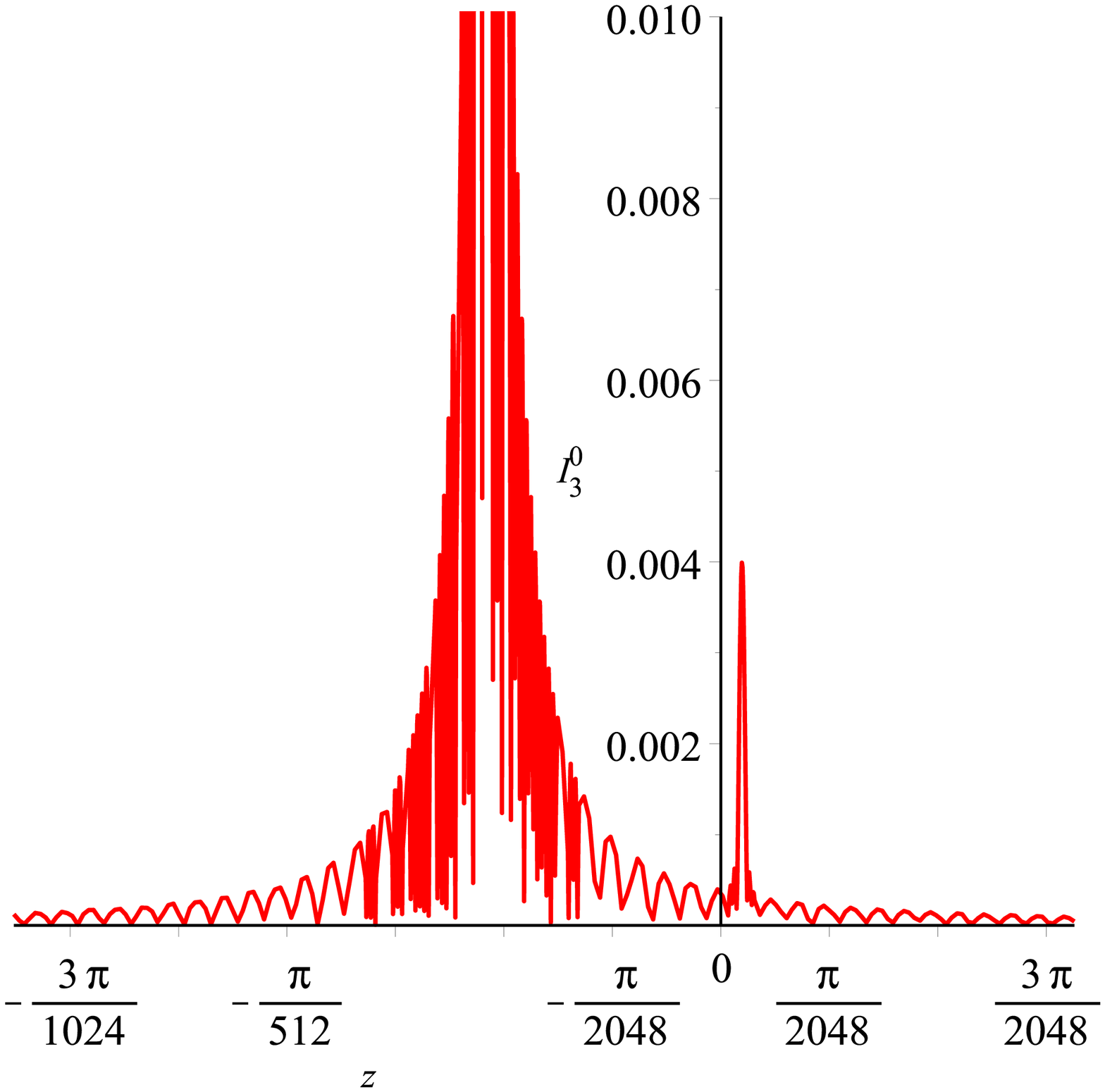}
\includegraphics[width=0.32\textwidth]{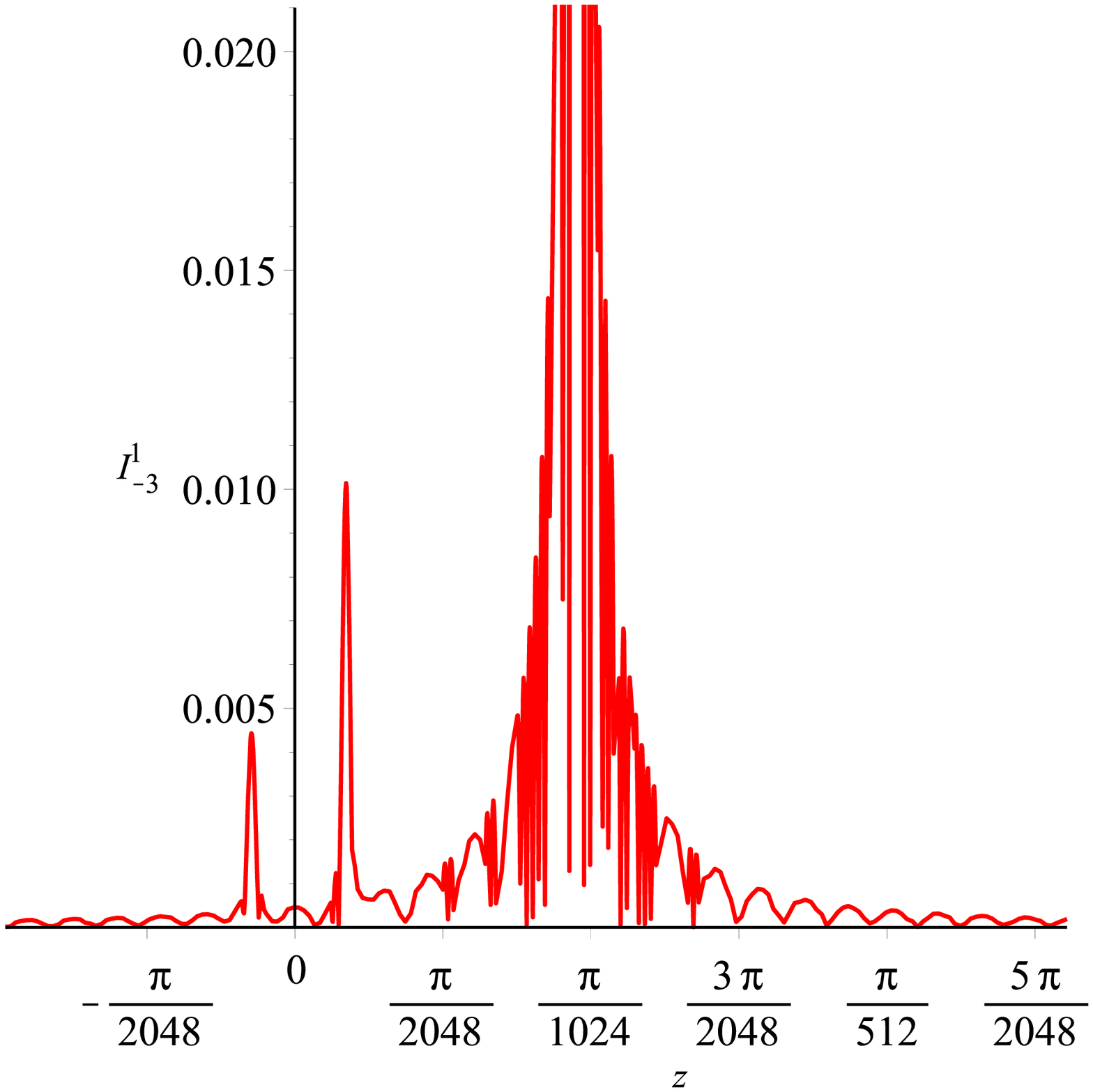}
\includegraphics[width=0.32\textwidth]{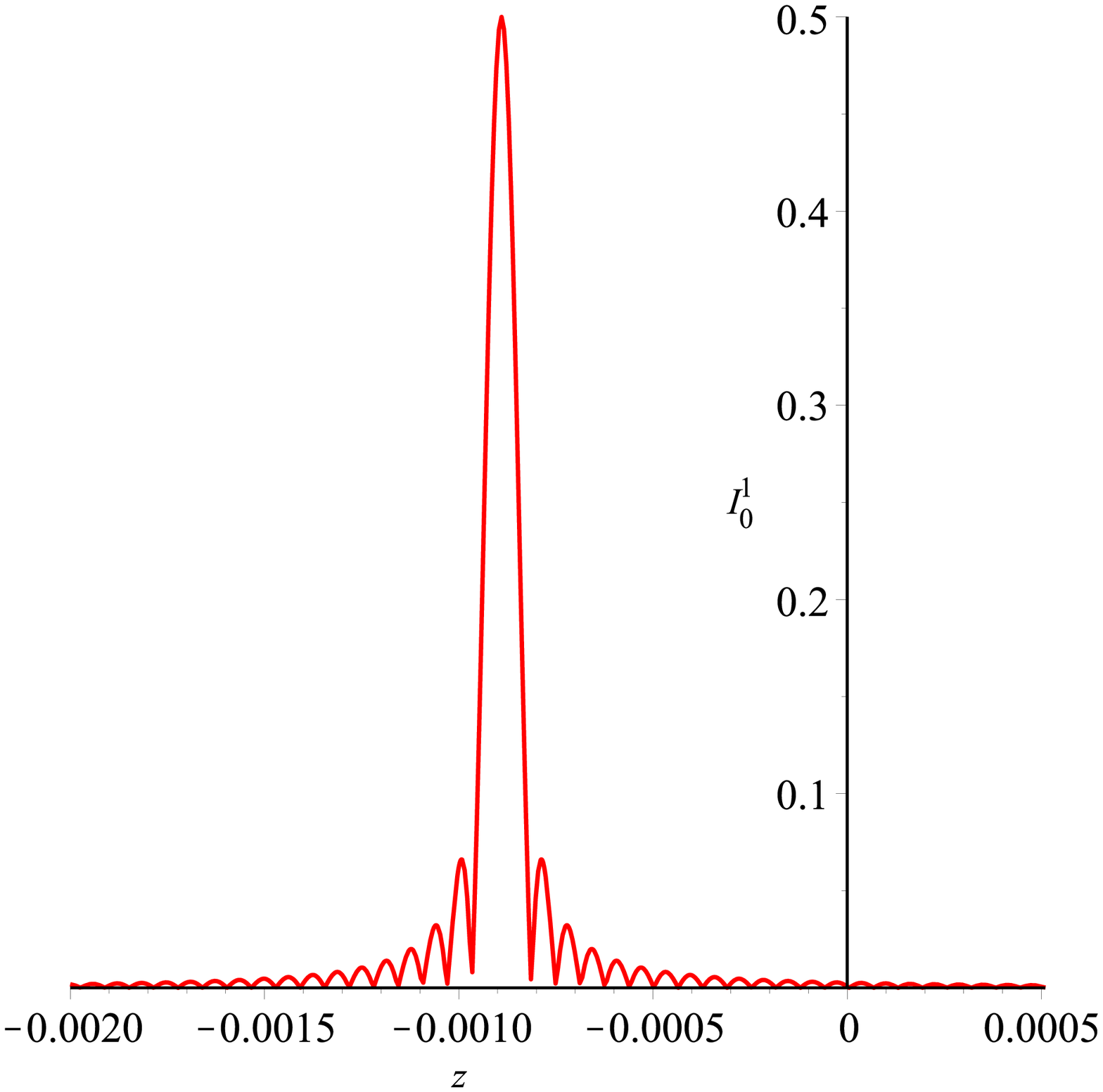}
\includegraphics[width=0.32\textwidth]{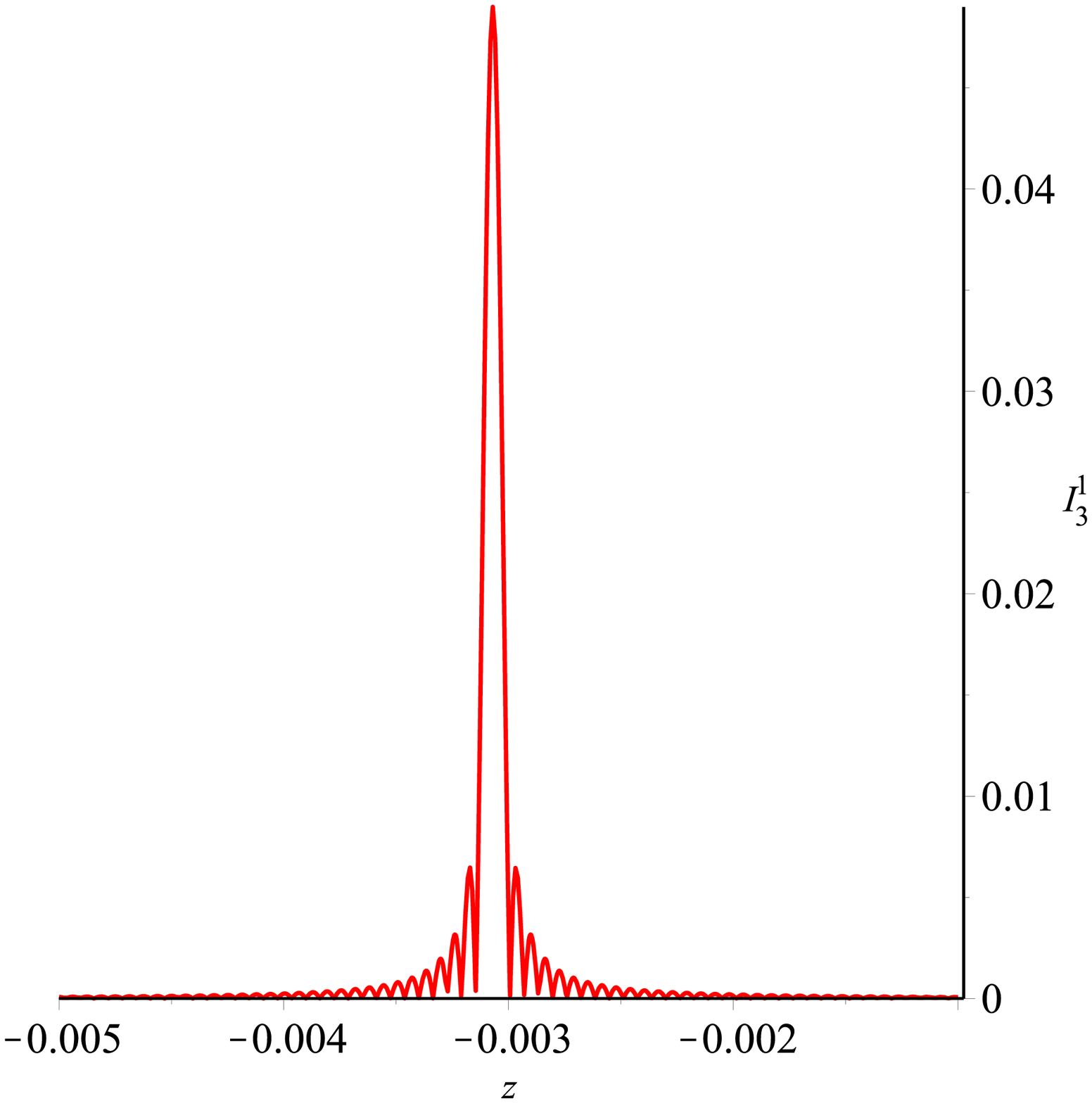}
\includegraphics[width=0.32\textwidth]{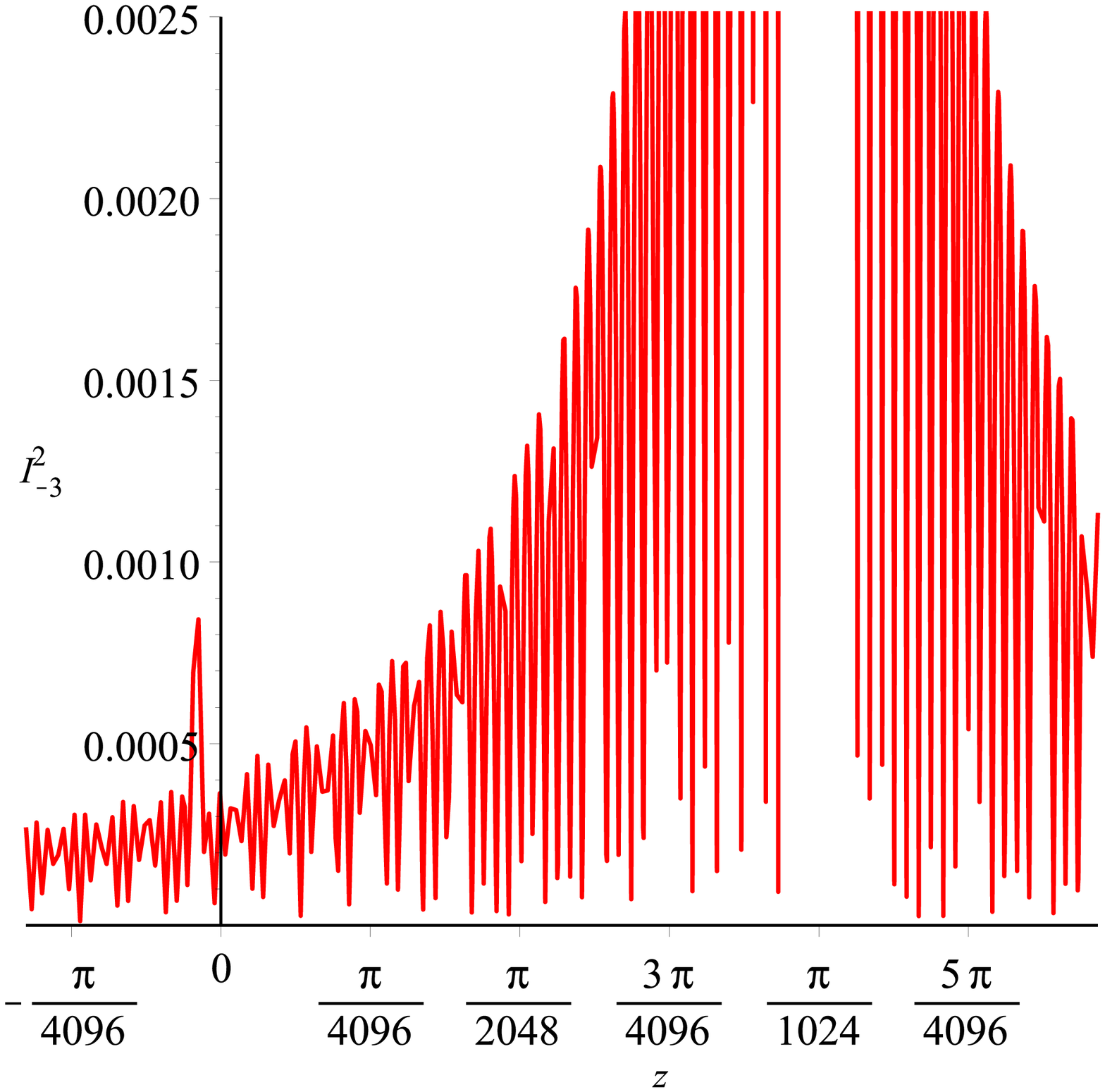}
\includegraphics[width=0.32\textwidth]{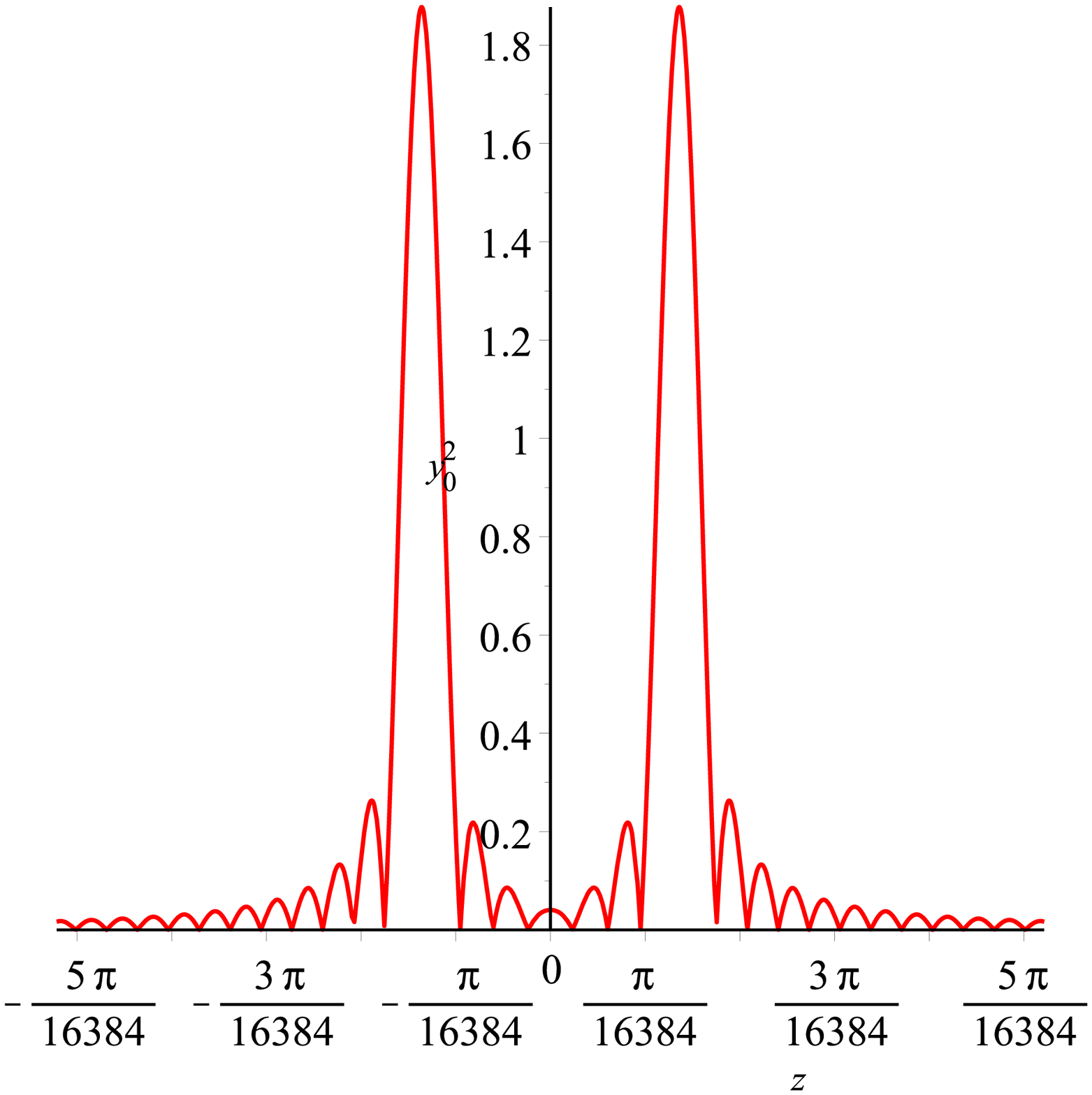}
\includegraphics[width=0.32\textwidth]{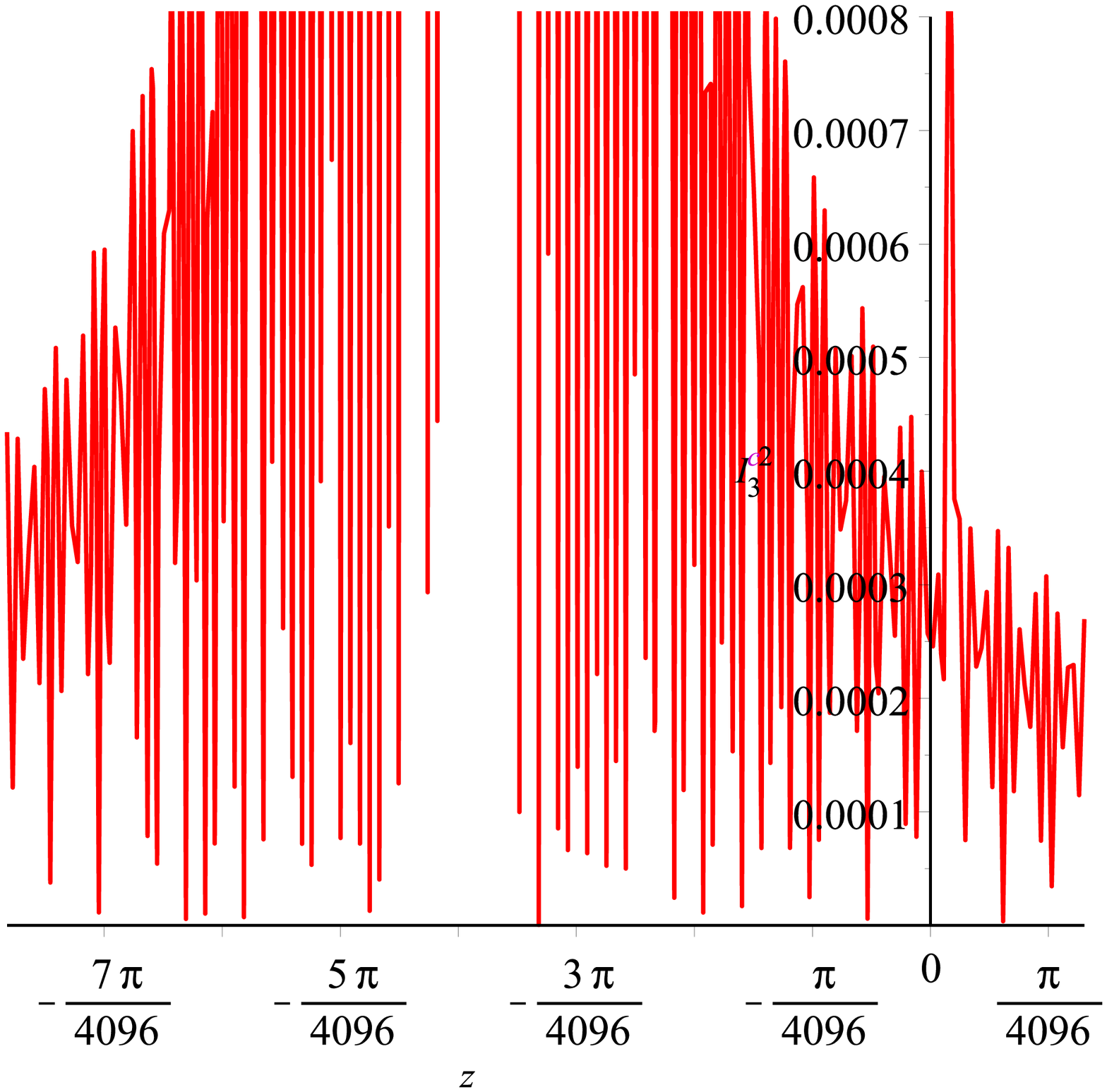}
\includegraphics[width=0.32\textwidth]{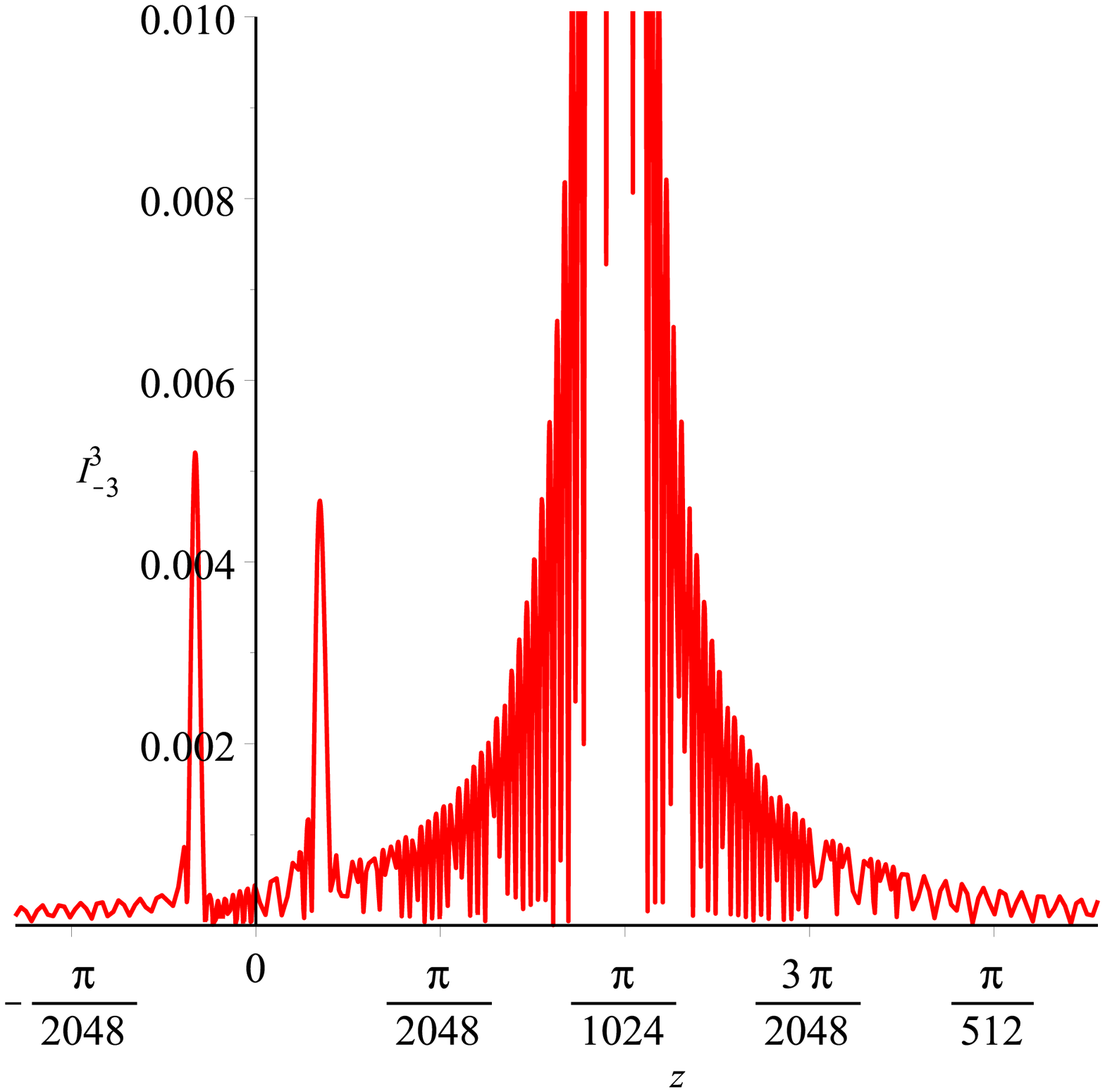}
\includegraphics[width=0.32\textwidth]{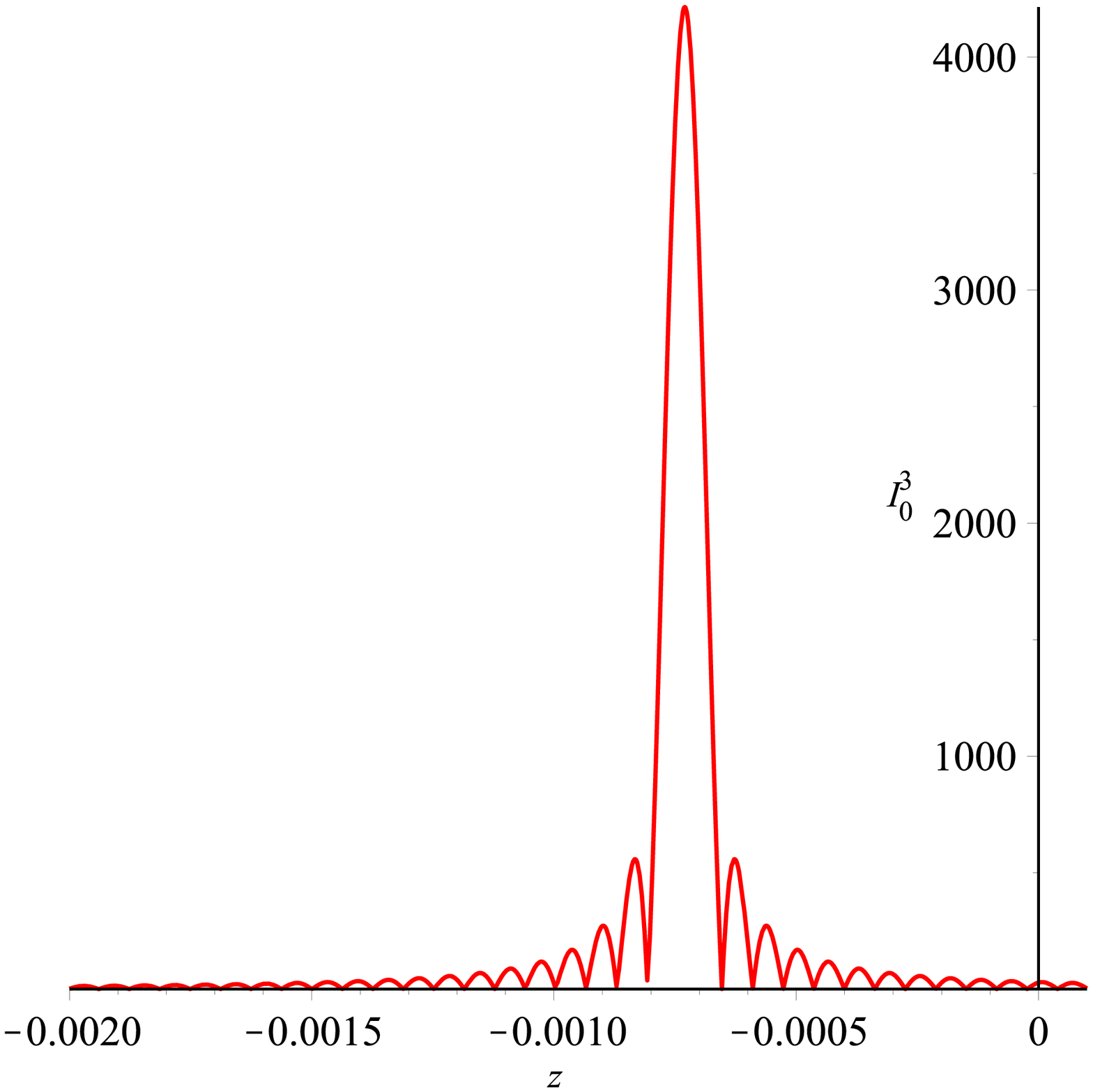}
\includegraphics[width=0.32\textwidth]{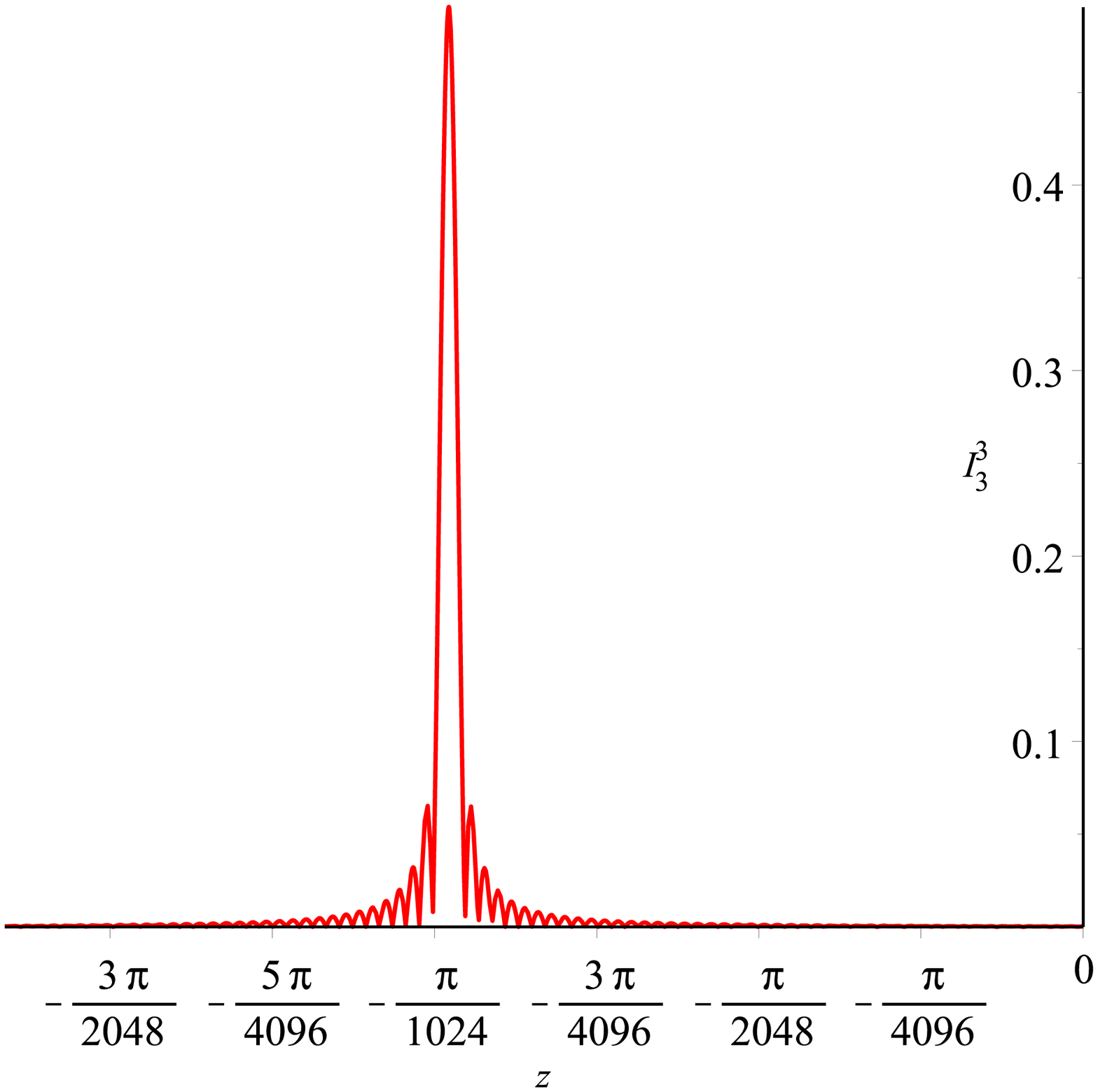}
} \caption{Diagrams of the waves intensity (\ref{intensityy}) for
point source positions  $y=0,\pm3$ for different lens potentials
$\delta=0,1,2,3$.  }
\end{figure}
\end{document}